\documentclass[aps,prb,twocolumn,groupedaddress,showpacs]{revtex4-1}
\usepackage{graphicx}% Include figure files
\usepackage{ulem}
\usepackage[english]{babel}
\usepackage[latin1]{inputenc}
\usepackage{subfigure}
\usepackage{amsxtra}
\usepackage{amsmath}
\usepackage{physics}
\usepackage{mathtools}
\usepackage{amssymb}
\usepackage{amstext,latexsym}
\usepackage{amsthm}
\usepackage{amsfonts}
\usepackage[rigidchapters,outermarks]{titlesec}
\usepackage{fancyhdr}
\usepackage{courier}
\usepackage{romannum}
\usepackage{txfonts,mathrsfs}
\usepackage{enumerate}
\usepackage[dvipsnames]{xcolor}

\begin{document}

\title{\textcolor{black}{Laser-driven quantum magnonics and THz dynamics of the order parameter in antiferromagnets}}

\author{D. Bossini} \email[]{davide.bossini@tu-dortmund.de}
\affiliation{Radboud University, Institute for Molecules and Materials, Heyendaalseweg 135, Nijmegen, The 
Netherlands}
\affiliation{Experimentelle Physik \Romannum{6}, Technische Universit{\"a}t Dortmund, D-44221 Dortmund, Germany}

\author{S. Dal Conte}
\affiliation{Dipartimento di Fisica, Politecnico di Milano, Piazza Leonardo da Vinci 32, Milano, Italy}
\affiliation{Istituto di Fotonica e Nanotecnologie, Consiglio Nazionale delle Ricerche, Piazza Leonardo da Vinci 32, Milano, Italy}

\author{O. Gomonay}
\affiliation{Institut f\"ur Physik, Johannes Gutenberg Universit\"at Mainz, D-55099 Mainz, Germany}
\affiliation{National Technical University of Ukraine ``KPI'', 03056, Kyiv, Ukraine}

\author{R. V. Pisarev}
\affiliation{Ioffe Physical-Technical Institute, Russian Academy of Sciences, 194021 St. Petersburg, 
Russia}

\author{M. Borovsak}
\affiliation{Jozef Stefan Institute \& CENN-Nanocenter, Jamova 39, Ljubljana SI-1000, Slovenia}

\author{D.Mihailovic}
\affiliation{Jozef Stefan Institute \& CENN-Nanocenter, Jamova 39, Ljubljana SI-1000, Slovenia}

\author{J. Sinova}
\affiliation{Institut f\"ur Physik, Johannes Gutenberg Universit\"at Mainz, D-55099 Mainz, Germany}
\affiliation{Institute of Physics ASCR, v.v.i., Cukrovarnicka 10, 162 53 Praha 6 Czech Republic}

\author{G. Cerullo}
\affiliation{Dipartimento di Fisica, Politecnico di Milano, Piazza Leonardo da Vinci 32, Milano, Italy}
\affiliation{Istituto di Fotonica e Nanotecnologie, Consiglio Nazionale delle Ricerche, Piazza Leonardo da Vinci 32, Milano, Italy} 

\author{J. H. Mentink}
\affiliation{Radboud University, Institute for Molecules and Materials, Heyendaalseweg 135, Nijmegen, The 
Netherlands}

\author{Th. Rasing}
\affiliation{Radboud University, Institute for Molecules and Materials, Heyendaalseweg 135, 
Nijmegen, The Netherlands}

\author{A. V. Kimel}
\affiliation{Radboud University, Institute for Molecules and Materials, Heyendaalseweg 135, 
Nijmegen, The Netherlands}

\date{\today}

\begin{abstract}
The impulsive generation of two-magnon modes in antiferromagnets by femtosecond optical pulses, so-called femto-nanomagnons, leads to coherent longitudinal oscillations of the antiferromagnetic order parameter that cannot be described by a thermodynamic Landau-Lifshitz approach. \textcolor{black}{We argue that this dynamics is triggered as a result of a laser-induced modification of the exchange interaction. In order to describe the oscillations} we have formulated a quantum mechanical description in terms of magnon pair operators and coherent states. 
\textcolor{black}{Such an approach allowed us to} derive an effective macroscopic equation of motion for the temporal evolution of the antiferromagnetic order parameter. An implication of the latter is
that the photo-induced spin dynamics represents a macroscopic entanglement of pairs of magnons with
femtosecond period and nanometer wavelength. By performing magneto-optical pump-probe experiments with 10 femtosecond resolution in the cubic KNiF$_3$ and the uniaxial K$_2$NiF$_4$ collinear Heisenberg antiferromagnets, we observed coherent oscillations at the frequency of 22 THz and 16 THz, respectively. The detected frequencies as a function of the temperature fit the two-magnon excitation up to the N\'eel point. The experimental signals are described as dynamics of magnetic linear \textcolor{black}{dichroism} due to longitudinal oscillations of the antiferromagnetic vector.
\end{abstract} 

\maketitle

\section{Introduction}
\label{section:Intro}

The research area of ultrafast \textcolor{black}{laser-induced} spin dynamics started two decades ago with the observation of sub-picosecond demagnetization of Ni by 60 femtosecond laser pulses~\cite{Beaurepaire1996} and the subsequent observation of the laser-induced ferromagnetic~\cite{vanKampen2002} and antiferromagnetic resonance~\cite{Kimel2004} in the time-domain\textcolor{black}{, triggered by laser-induced heating and optically generated effective magnetic field~\cite{Kimel2005a,Satoh2010,Nemec2013,Stanciu2007}. These experiments opened up new opportunities for the generation and the control of propagating spin waves with sub-picosecond temporal resolution~\cite{Satoh2012,Hashi2017,Hashi2018}. It even ignited a surge of interest in the field of photo-magnonics promising to develop magnon-based information processing into the THz domain~\cite{Lenk2011}.}  

On the fundamental side, driving spins out of equilibrium with femtosecond laser pulses is expected to launch dynamics beyond the realm of classical mechanics and thermodynamics~\cite{Li2013}. Nevertheless, all the manifestations of light-induced (sub)-picosecond spin dynamics have been hitherto interpreted by means of classical equations of motion.~\cite{Kimel2005a,Bossini2014,Kalashnikova2008,Satoh2010,Kalashnikova2007,Radu2011} \textcolor{black}{This approach was proven to be} successful if the photo-generated single-magnon modes have wavevectors near the center of the Brillouin zone.

\begin{figure}[t!]
	\center
	\includegraphics[]{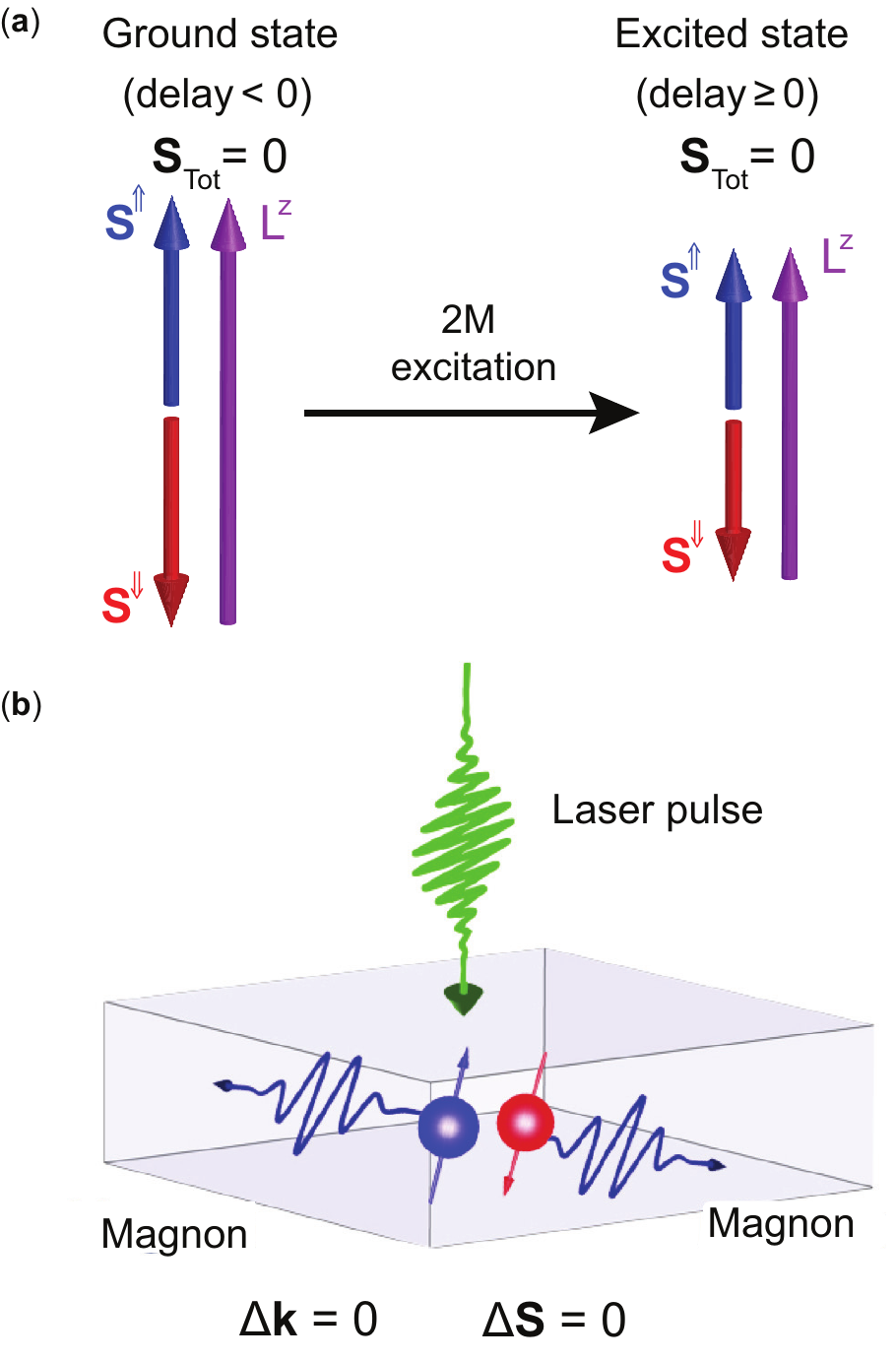}
	\caption{\footnotesize{Conservation laws of the ISRS excitation of the 2M-mode. (a) The total spin is conserved, since magnons are generated by spin flip events occurring in opposite sublattices. (b) The visible excitation light pulse has almost zero wavevector, thus only pairs of magnons with the same and opposite wavevectors can be triggered. As a result, the total wavevector exchanged in the process is zero.}}
	\label{fig:2MgenerationHelen}
	\vspace{-0.5cm}	
	\end{figure}

It is well known that non-zero wavevector magnons can be optically excited via two-magnon (2M) \textcolor{black}{processes. Obeying the laws of conservation of energy and momentum a photon with the energy $\hbar\omega_{p1}$ and momentum $k_{p1}$ can generate two magnons with energies $\hbar\omega_{m1}$, $\hbar\omega_{m2}$ and momenta $k_{m1}$, $k_{m2}$ via the Raman scattering process. As a result, the photon is scattered with the energy $\hbar\omega_{p2}$ and momentum  $k_{p2}$ so that $\hbar\omega_{p1} = \hbar\omega_{m1} + \hbar\omega_{m2} + \hbar\omega_{p2}$ and $k_{p1} = k_{m1} + k_{m2} + k_{p2}$. For visible light and magnons far away from the center of the Brillouin zone, it can be safely assumed that $k_{m1,2} \gg k_{p1,2}$. Consequently, light generates two counter-propagating magnons $k_{m1} \approx k_{m2}$ and $\omega_{m1} \approx \omega_{m2}$.}
 \textcolor{black}{The dominating light-matter process is the interaction of the electric field of light with electric charges; according to the selection rules of electric dipole transitions the total spin in the excitation of the 2M mode is conserved (see Fig. \ref{fig:2MgenerationHelen}). An effective generation of magnon pairs at the edges of the Brillouin zone, where the magnon density of states is the largest, was demonstrated via spontaneous Raman (SR) scattering.~\cite{Fleury1968,Chinn1971,Fleury1970,Balucani1973,Cottam1986,Lemmens2003}
Femtosecond laser pulses and the mechanism of impulsive stimulated Raman scattering (ISRS)~\cite{Bossini2016,Bossini:2017bo,Zhao2004,Zhao2006} led to the generation of pairs of magnons and to the observation of the subsequent spin dynamics in time-domain with temporal resolution on the order of 10 femtoseconds. 
Unlike all the previous studies, the first time-resolved two-magnon experiments allowed to claim that the triggered spin dynamics cannot be understood in the frame of classical physics.~\cite{Zhao2004,Zhao2006} It was reported that the generation of the dynamics of correlations involving pairs of spins  in the antiferromagnetic insulator MnF$_2$ induces squeezing. The spin fluctuations in this squeezed state vary periodically in time and are reduced below the level of the ground-state quantum noise. More recently, Bossini et al. reported that the photo-excitation of pairs of magnons with wavevector near the edges of the Brillouin zone, named \textit{femto-nanomagnons},~\cite{Bossini2016} in antiferromagnetic KNiF$_3$ triggers dynamics of the antiferromagnetic order parameter $\mathbf{L}$. This quantity is defined in terms of the magnetizations of the two magnetic sublattices ($\mathbf{M}_1$,$\mathbf{M}_2$) $\mathbf{L} \equiv \mathbf{M}_1- \mathbf{M}_2$.~\cite{Bossini2016} The generation of pairs of magnons does not simply reduce the magnitude of the order parameter, but it triggers longitudinal oscillations of the modulus of $\mathbf{L}$ at the frequency of the 2M excitation ($2\omega_{m}$).
Despite the highly intriguing results, employing the quantum regime of spin dynamics in photo-magnonics is prevented by poor understanding of the fundamental physics of the process. The experimental observations reported in Ref.~\cite{Zhao2004,Zhao2006,Bossini2016} have not found an explanation yet or even contradicted what has been reported before.}

First of all,  while the detection of two-magnon dynamics in Ref.~\cite{Zhao2004,Zhao2006} was based on the time-resolved measurement of the transmissivity, Ref.~\cite{Bossini2016} employed time-resolved measurements of the polarization rotation originating from antiferromagnetic linear dichroism.  It is not clear if these two detection schemes will give similar results: can the length oscillations of $\mathbf{L}$ reported in Ref.~\cite{Bossini2016} be interpreted in terms of the squeezed magnon states from Ref.~\cite{Zhao2004,Zhao2006} or do Ref. ~\cite{Zhao2004,Zhao2006, Bossini2016} report mutually independent phenomena? \textcolor{black}{The magneto-optical experiment~\cite{Bossini2016} demonstrated the possibility to control the phase of the oscillations of the magneto-optical signal varying the polarization of the exciting laser pulse. However the theory behind this process remains unclear, since it is not established whether the observed modification of the magneto-optical signal depends on a change of sign of the oscillations of $\mathbf{L}$ or on a photo-induced modification of the magneto-optical tensor.}
Based on the temperature dependence of the efficiencies of the stimulated 2-magnon Raman scattering in FeF$_2$, it was suggested~\cite{Zhao2006} that contrary to the spontaneous Raman process long-range spin order is an important, if not essential, component of the coherent two-magnon scattering. It is not clear if this may be a feature specific to FeF$_2$ or a general phenomenon relevant to all antiferromagnets. Aiming to clarify these questions, this work focuses on theoretical and experimental studies of the impulsive stimulated Raman scattering and subsequent spin dynamics.
 
In particular, we show that the description of spin dynamics triggered by the generation of pairs of femto-nanomagnons can be simplified by introducing magnon pair operators. Using coherent states for these operators, we are able to derive an effective macroscopic description beyond the conventional Landau-Lifshitz phenomenology. Moreover, the commonly employed concept of light-induced effective field to describe the photo-excitation of macrospin dynamics~\cite{Kirilyuk2010,Kalashnikova2007,Kalashnikova2008} does not apply to the femto-nanomagnons. In our theoretical framework we formulate an analogous concept, a generalized force responsible for the spin dynamics. 

Experimentally the temperature-dependence and the pump-polarization dependence of the spin dynamics was explored after impulsive generation of femto-nanomagnons in two antiferromagnets KNiF$_3$ and K$_2$NiF$_4$ having different magnetocrystalline anisotropy. The microscopic theory does not predict a modification of the spin dynamics if the polarization of the pump beam is changed. In agreement with the theoretical predictions no polarization dependence of the amplitude and phase of the oscillations of the magneto-optical signal was observed in K$_2$NiF$_4$. 
However, a dependence was clearly observed in KNiF$_3$. A phenomenological analysis reveals that the perturbations of the spin system induced by different polarization states are not equivalent, although a quantitative description of the experimental observation is still elusive. Using such phenomenological approach we suggest a possible origin of this phenomenon.  

This paper is organized as follows. In Sec. \ref{section:Theory} the quantum mechanical theory describing the spin dynamics initiated by the generation of femto-nanomagnons is reported. Section \ref{section:M&M} describes the experimental techniques, together with the properties of the materials investigated. Section \ref{section:Tdep} reports the investigation of the temperature dependence of the femto-nanomagnonic dynamics. The experimentally verified criterion allowing to select the proper conditions for the phase control is discussed in Sec. \ref{section:PumpPolDependence}. The conclusions and perspectives of our work are reported in Sec. \ref{section:conclusions}.

\section{Theory}
\label{section:Theory}

\textcolor{black}{In this section we present a theoretical description of spin dynamics induced by the two-magnon mode ("2M" in the following), meaning with this expression a pair of magnons with the same frequency and same wavevector in magnitude, but opposite in sign. First, we provide an exclusively qualitative discussion of 2M dynamics, highlighting the qualitative differences between a classical and quantum description. The results of our entire modelling are here summarised and reported without the mathematical formalism, which is then employed in the rest of the section. Second, we introduce a novel microscopic quantum description of 2M dynamics in terms of boson-pair operators. They allow for a simple analytical treatment, both at zero- and at finite-temperature. In the third part we show that using coherent states for the boson-pair operators, we can derive an effectively macroscopic theory for the longitudinal dynamics of the antiferromagnetic vector, which supplements the phenomenological Landau-Lifshitz description for spin dynamics on the femtosecond timescale. Moreover, within this macroscopic description we are able to define a generalized force, analogous to the (light-induced) effective magnetic field, commonly employed for long-wavelength magnons. Fourth, we analyze in detail the polarization-dependence using a phenomenological treatment of light-matter interaction and we compare this to the results obtained from the quantum model. Finally, we elaborate on various quantum aspects of 2M dynamics and demonstrate that a natural and unavoidable implication of our theory is that the photo-induced spin dynamics drives entanglement of magnon pairs and, therefore, is a genuine quantum effect.}

\subsection{Qualitative description of 2M dynamics}\label{secqualitative}
\textcolor{black}{Two-magnon dynamics has been extensively discussed in the frequency domain, mainly in the context of spontaneous Raman scattering \cite{Fleury1968,Chinn1971,Fleury1970,Balucani1973,Cottam1986,Lemmens2003}. While these theoretical descriptions are essentially quantum, it is not completely clear whether a quantum description is strictly necessary, or arguments in terms of classical spin waves would be adequate as well. Here we are interested in a description of 2M dynamics in the time domain, triggered by an impulsive stimulated Raman scattering process. To investigate the need for a quantum treatment, we first analyze 2M dynamics with classical spin wave theory and outline the qualitative features. Second, we show that a qualitatively distinct dynamics arises when quantum correlations between spins at different positions are taken into account. We explain why such quantum features are measurable in macroscopic systems at elevated temperature and elaborate on the excitation mechanism. Finally, we argue that the quantum treatment is required to capture the dynamics observed in the experiments presented in Secs.~\ref{section:Tdep}-\ref{section:PumpPolDependence}.}

%\begin{figure}[h]
%	\centering
%	\includegraphics[width=1.0\columnwidth]{DispersionModes.eps}
%	\caption[schema]{ Difference between low-energy magnons and femto-nanomagnons. The central panel shows the dispersion relation calculated for KNiF$_3$. The upper and lower panels show schematically the dynamical properties of the typical eigenmodes with corresponding $\mathbf{k}$ framed in the dispersion curve. Magnons with $\mathbf{k}\approx 0$ correspond to excitations of both magnetic sublattices with almost antiparallel orientation of spins. Femto-nanomagnons with $\mathbf{k}\approx \mathbf{k}_\mathrm{R}$ correspond to excitation of mainly one of the two magnetic sublattices and could be classified according to the direction of the quantization axis as $\uparrow$ and $\downarrow$ modes.}
%	\label{Fig:fig_dispersion_modes}
%\end{figure}

\textcolor{black}{\subsection{Classical description of two-magnon dynamics and its limitations.}} At long wavelengths, the dynamics of magnons in antiferromagnets is conveniently described by the Landau-Lifsthitz-type equations of motion for the sublattice magnetizations $\boldsymbol{S}^{\Uparrow}$ and $\boldsymbol{S}^{\Downarrow}$, which are defined as thermodynamic averages of the local spins over physically small volumes (so-called mean field approximation). This classical antiferromagnetic state is usually described in terms of two macroscopic vectors, the magnetization $\mathbf{M}$ and the N\'eel vector $\mathbf{L}$, the latter is canonically introduced as order parameter for an antiferromagnet:~\cite{Landau1984}
		\begin{equation}
	%		\begin{array}
\mathbf{M}= \frac{N}{2}\left( \boldsymbol{S}^{\Uparrow} +\boldsymbol{S}^{\Downarrow}\right),
\quad
\mathbf{L} = \frac{N}{2}\left( \boldsymbol{S}^{\Uparrow} - \boldsymbol{S}^{\Downarrow}\right),
	%		\end{array}
	\label{eq:DefIntro1}
	\end{equation}
\noindent where $N$ is the number of magnetic atoms ($N/2$ per sublattice) per unit volume. These definitions hold for two-sublattice antiferromagnets. For the sake of simplicity we assume that $\hbar = 1$ and that the gyromagnetic ratio equals 1 as well in the definition of both $\mathbf{M}$ and $\mathbf{L}$. Within this classical picture the dynamics of the N\'eel vector results in the emergence of a small but nonzero magnetization $\mathbf{M}\sim\mathbf{L}\times\partial_t\mathbf{L}$ (so-called dynamic magnetization). Hence, both magnetic sublattices are involved in the homogenous oscillation. If the equilibrium orientation of the N\'eel vector is along the $z$-axis, oscillations of $\mathbf{L}$ at the frequency of antiferromagnetic resonance $\omega_{\mathrm{AFM}}$ can be launched by inducing transverse components along the $x$- and $y$-axes. In terms of magnons, the frequency $\omega_{\mathrm{AFM}}$ is the eigenfrequency of spin waves at the center of the Brillouin zone. In other words, the excitation of oscillations of $\mathbf{L}$ corresponds to injection of coherent magnons with approximately zero-wavevector. As long as the magnons retain mutual coherence, the modulus of the N\'eel vector is not reduced, transverse components ($x$ and $y$) oscillate at the frequency $\omega_{\mathrm{AFM}}$ and the $z$-component at the frequency $2\omega_{\mathrm{AFM}}$. If the magnons were injected via a torque induced by a resonant magnetic-field at the frequency $\omega_{\mathrm{AFM}}$, the amplitude of the oscillations of the transverse components would scale linearly with the magnetic field H. On the other hand the amplitude of the $2\omega_{\mathrm{AFM}}$ oscillations of the $z$-component would be proportional to H$^2$, as reported~\cite{Baierl2016}. If the magnons were triggered via ISRS, the amplitude of the transverse and $z$-components would scale linearly and quadratically with respect to the intensity of the pump beam, respectively.

\begin{figure}
\centering
\includegraphics[width=1.0\columnwidth]{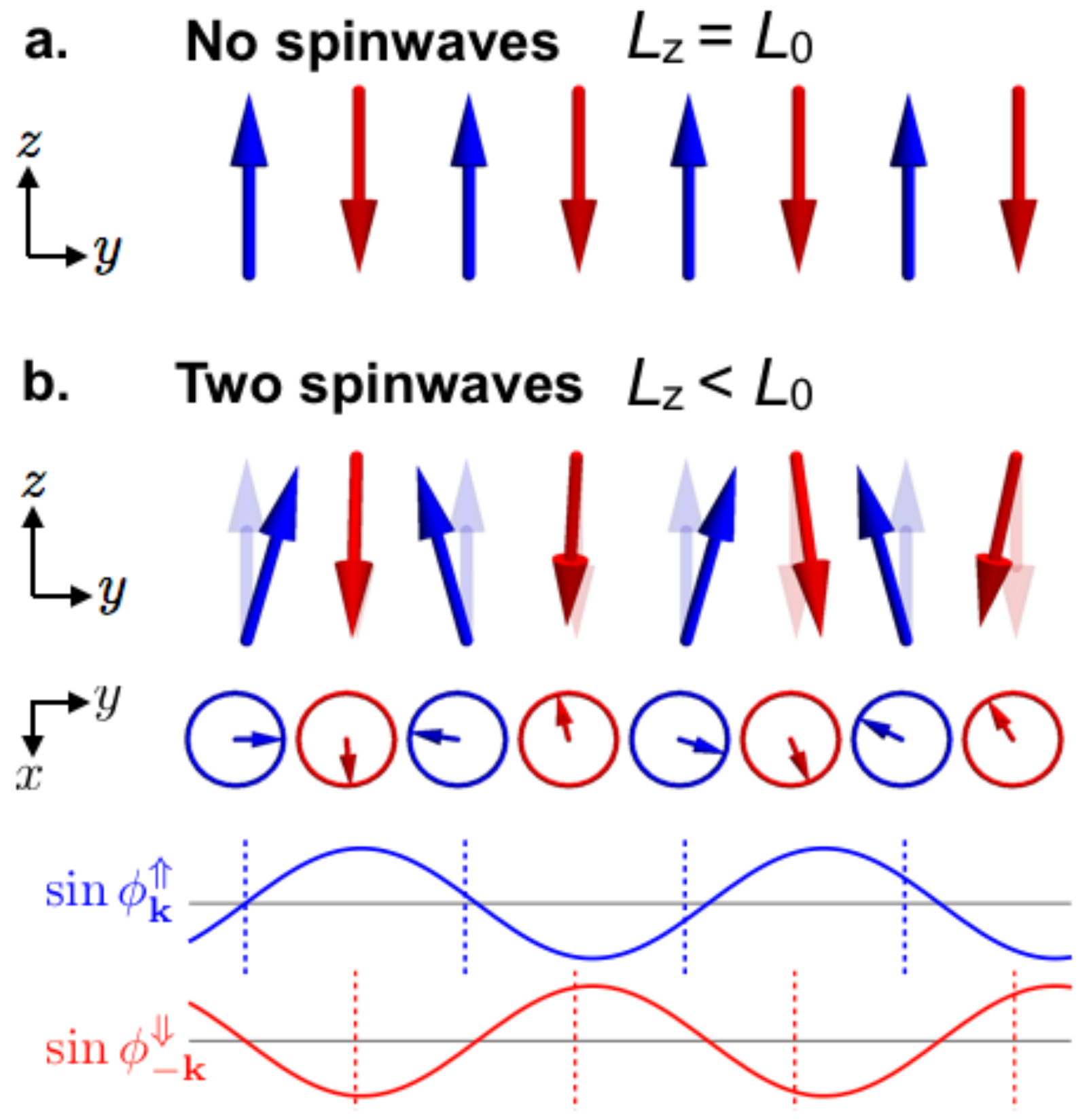}
\caption{ {\bf Illustration of 2M excitation using classical antiferromagnetic spin waves}. a) classical N\'eel state with antiparallel spins at adjacent sites. \textcolor{black}{b) State with two spinwaves, one in each sublattice with opposite wavevector $k = \pm 0.95$ $\pi/2a$ close to the Brillouin zone boundary, where $a$ is the lattice constant. Since in each sublattice the spin is reduced by one unit, the total magnetization is conserved but the N\'eel vector is reduced. The two bottom panels show the corresponding spin projections in the transverse plane and their phase relationship, with dashed lines indicating lattice points. Spins within the same sublattice are nearly out of phase, while spins in different sublattices are nearly perpendicular in the transverse plane. Therefore, the torques induced by neighboring spins cancel almost completely out and the spin waves can be considered localized in one sublattice.}}\label{twomagnon}
\end{figure}

\textcolor{black}{Classical spin wave theory is nevertheless not restricted to homogenous $\mathbf{k}\approx 0$ oscillations. In the framework of an atomistic picture, in which the spin states of individual atoms are disentangled from each other, we consider spin waves with any wavevector in the magnetic Brillouin zone. In particular a magnetic excitation triggered by light, with almost-zero wavevector, can consist of a pair of magnons belonging to different modes, i.e. with wavevector $\mathbf{k}$ and $-\mathbf{k}$ and with the same frequency. Relevant to our case, for $\mathbf{k}$ close to the Brillouin zone boundary, spinwaves are almost localized on one magnetic sublattice. \textcolor{black}{This peculiarity is due to the short-wavelength, which is comparable with the lattice constant.} Hence, we can envision exciting two distinct spinwaves, each one perturbing one of the magnetic sublattices (for simplicity we assume here that the modes are strictly localized, but the argument also holds for eigenmodes that are themselves superpositions of spinwaves in each of the sublattices). An illustration of this scenario is given in Fig~\ref{twomagnon}. Similar to the $\mathbf{k}\approx 0$ case, the local spin oscillations are, to leading order, transverse to the equilibrium value of $\mathbf{L}$, with a well-defined phase relation between spin deflections at different positions (see Fig~\ref{twomagnon}). Although two spinwaves are excited, the oscillation frequency for these transverse deflections is the frequency of each single spin wave mode. Within linear spin wave theory the net change of the longitudinal magnetization is zero, since the change of the local magnetization in each of the magnetic sublattices has opposite sign. On the contrary, the length of the N\'eel vector is reduced as compared to the equilibrium value. Analogously to the spin waves at the center of the Brillouin zone, also in this case the second harmonic can appear in the $z$-component as the next-to-the-leading-order dynamic contribution which scales quadratically with fluence.}

\vspace{0.3cm}
\subsection{Quantum description of 2M dynamics.} From the aforedescribed analysis it follows that within classical spin wave theory the leading order dynamics is transverse to the equilibrium direction of the N\'eel vector, while the longitudinal dynamics of the N\'eel vector occurs at the next-to-the-leading order, at the double frequency of the transverse oscillations and with amplitude scaling quadratically with the excitation fluence. In the following we will analyze the situation in which quantum correlations between the spin states of the neighbouring magnetic atoms cannot be neglected, meaning that we cannot rely on the mean-field approach. For simplicity, we start with a simple example of just two quantum spins with $S=1/2$. Quantum correlations $\langle\hat{S}_1\hat{S}_2\rangle\neq\langle\hat{S}_1\rangle\langle\hat{S}_2\rangle$ reveal themselves for example when the system is in the superposition state: $|\psi\rangle = c\,|\!\uparrow_1\rangle~\!|\!\downarrow_2~\!\rangle + d\,|\!\downarrow_1~\!\rangle|\!\uparrow_2~\!\rangle$, where the symbols $\uparrow$ and $\downarrow$ indicate two different spin orientations. For this state, the total spin $\langle\hat{\mathbf{S}}_1+\hat{\mathbf{S}}_2\rangle=0$, but this does not exclude variation of the individual components $\langle\hat{\mathbf{S}}_1\rangle$, $\langle\hat{\mathbf{S}}_2\rangle$ (where the brackets mean quantum mechanical average) which means that the length of the N\'eel vector defined as $|\mathbf{L}|\equiv|\langle\hat{\mathbf{S}}_1-\hat{\mathbf{S}}_2\rangle |=|c|^2-|d|^2$ can vary. Such changes can be viewed as an elongation of one spin correlated with a shrinking of the other spin, which is accompanied by changes in the spin correlations $\langle\hat{S}^z_1\hat{S}^z_2\rangle=-(|c|^2+|d|^2)/4$ and $\langle\hat{S}^x_1\hat{S}^x_2\rangle=\langle\hat{S}^y_1\hat{S}^y_2\rangle=(cd^*+c^*d)/4$.

\textcolor{black}{{In a more general picture of an antiferromagnet with $N$ correlated magnetic atoms, the magnetization and the N\'eel vector are defined through both quantum-mechanical average and average in real space ($i$ and $j$ are indices for lattice sites):
	\begin{align}
%		\begin{array}
&\mathbf{M} \coloneqq \left(\sum_{i} \langle \hat{\mathbf{S}}^{\Uparrow}_{i}\rangle + \sum_{j}\langle \hat{\mathbf{S}}^{\Downarrow}_{j}\rangle\right), \nonumber\\
&\mathbf{L}  \coloneqq \left(\sum_{i} \langle \hat{\mathbf{S}}^{\Uparrow}_{i}\rangle - \sum_{j}\langle\hat{\mathbf{S}}^{\Downarrow}_{j}\rangle\right) .
%		\end{array}
\label{eq:DefIntro}
\end{align}	
and the sums are considered in the unit volume. In the limit of vanishingly small correlations between neighboring spins definition Eq.~(\ref{eq:DefIntro}) coincides with the classical vectors Eq.~(\ref{eq:DefIntro1}).}}

\begin{figure}[hb]
\centering
\includegraphics[width=1.0\columnwidth]{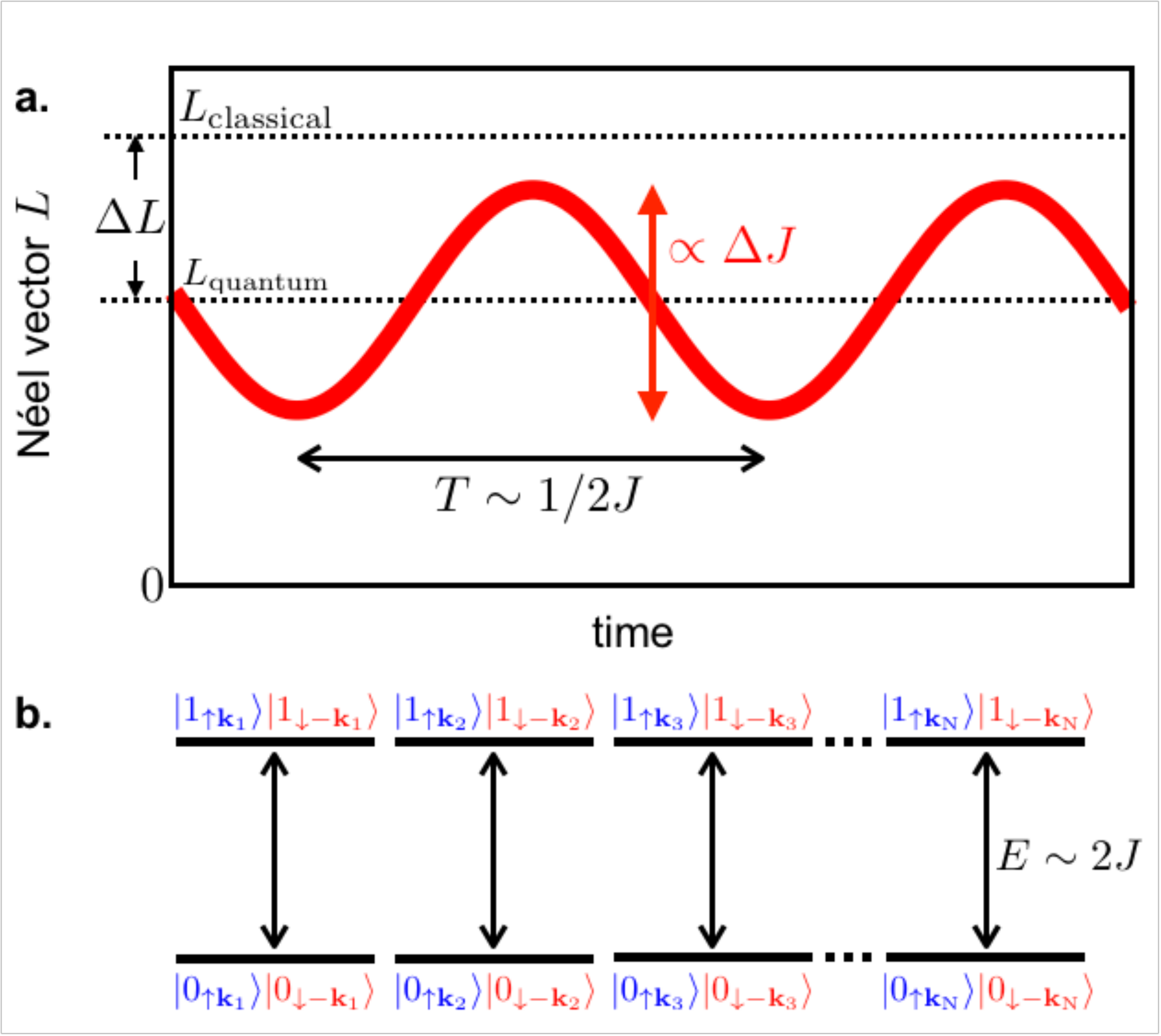}
\caption{ {\bf Illustration of the oscillation of the N\'eel vector}. a) In a quantum antiferromagnet, the magnitude of the N\'eel vector $L=|\mathbf{L}|$ is reduced with respect to the classical value by an amount $\Delta L$ (dashed lines). This reduction originates from dressing with two-magnon excitations. An ultrafast perturbation of the exchange interaction ($\Delta J$) changes the contribution of these dressed states and triggers longitudinal oscillations. b) The emergence of such oscillations can be understood from the coherent excitation of an ensemble of two-level systems. Each of these two level system represents coherent oscillations between the N\'eel state ${|0_{\uparrow\mathbf{k}_i}\rangle}{|0_{\downarrow-\mathbf{k}_i}\rangle}$ and a state with two magnons excited ${|1_{\uparrow\mathbf{k}_i}\rangle}{|1_{\downarrow-\mathbf{k}_i}\rangle}$, seperated by an energy $E\sim 2J$.}\label{Losc}
\end{figure}

\textcolor{black}{In such extended systems, quantum correlations reveal themselves in a coherent dynamics which can be described as quantum Rabi-like oscillations between the ground N\'eel state and the excited state (i.e. 2M state), both of which are represented in Fig.~\ref{twomagnon}. Since the magnitude of $\mathbf{L}$ is reduced in the state with 2M as compared to the N\'eel state, a time-dependent superposition of these two states gives rise to longitudinal oscillations of $\mathbf{L}$, already within the harmonic magnon theory. Hence, we can understand the 2M dynamics relying on a simplified picture of a two-level system, in which coherent quantum oscillations occur between two-particle states: the N\'eel state, which can be expressed in terms of absence of any magnons i.e. ${|0_{\uparrow\mathbf{k}}\rangle}{|0_{\downarrow-\mathbf{k}}\rangle}$, and a state in which the 2M-mode is excited ${|1_{\uparrow\mathbf{k}}\rangle}{|1_{\downarrow-\mathbf{k}}\rangle}$ (see Fig.~\ref{Losc}). An extended antiferromagnetic systems can be envisaged as a large ensemble of such two-level systems, one for each $\mathbf{k}$-value. A short optical excitation pulse in the ISRS scheme triggers oscillations with the same initial phase for each two-level system. As long as the two-level systems remain mutually coherent, the length of the $\mathbf{L}$ vector oscillates with a frequency twice bigger than the coherent magnons. Although the spectrum of magnons in a bulk antiferromagnet is broad, the overwhelming contribution to the magnon density of states originates from magnons with wavevector close to the Brillouin zone boundary, $\mathbf{k}\approx\mathbf{k}_\text{R}=[\pm\pi/2a, \pm\pi/2a,\pm\pi/2a]$, where $a$ is the lattice constant. For such magnons, the energy of a single quantum is $\hbar \omega_\text{R}\approx zSJ$, where $z$ is the number of nearest neighbour spins with spin $S$ and $J$ is the parameter of the exchange interaction. It means that as long as magnons with energy $\hbar \omega_\text{R}$ and opposite $\mathbf{k}$ remain coherent, the N\'eel vector oscillates at the frequency $2\omega_\text{R}$. Hence, within the quantum description the double frequency is already predicted within the harmonic magnon theory. This is the main difference between the quantum and classical descriptions. In particular, if the N\'eel vector is aligned along the $z$-axis and pairs of coherent magnons with equal frequencies and opposite wavevectors are photo-induced via ISRS, the quantum theory predicts oscillations of the length of the antiferromagnetic vector at a frequency which is twice the frequency of the individual magnons. The amplitude of the oscillations will scale linearly with the pump fluence. In the classical theory the length of the N\'eel vector does not oscillate in a linear regime, although oscillations of the $z$-component of the N\'eel vector can be achieved, in which case the amplitude of the oscillations scales quadratically withe the pump fluence. In Sec.~\ref{subsection:microscopictheory} below we focus exclusively on a quantum description of 2M dynamics.}

\subsection{Excitation mechanism}\label{subsecexcimech} \textcolor{black}{Next we elaborate on the excitation mechanism in the quantum mechanical framework. For an individual two-level system, the energy splitting is given by $E=2\hbar\omega_\text{R}=2zJS$. Although a classical antiferromagnet can be described by the N\'eel state, this is not an eigenstate of the Heisenberg Hamiltonian. Therefore the ground state of a quantum antiferromagnet must be different; in particular it is given by the superposition of the N\'eel state and states in which two correlated spin-flips and thus magnons are excited. The correlated spin-flips are induced by the canonical ladder operators  $\hat{S}_i^+\hat{S}_j^-$ appearing in the Heisenberg Hamiltonian and represent transitions $\cdots|\!\uparrow\rangle|\!\downarrow\rangle|\!\uparrow\rangle|\!\downarrow\rangle\cdots \rightarrow \cdots|\!\uparrow\rangle|\!\uparrow\rangle|\!\downarrow\rangle|\!\downarrow\rangle\cdots$. This causes a ground state in which the N\'eel vector is reduced with respect to the classical value, as illustrated by dashed lines in Fig.~\ref{Losc}. These fluctuations, which are a purely quantum mechanical effects, are not thermal and thus are present even at zero temperature. Since the fluctuations are dominated by long-wavelength low-energy magnons, they are suppressed as the temperature increases. Moreover, no phase relation exists among the magnons generated in this process.}

\textcolor{black}{The initial state thus has a non-zero population of the magnon states relevant for the 2M-mode. Thus a sudden perturbation of the exchange interaction $\Delta J$ is sufficient to induce coherent oscillations of the population between the ground state and the excited 2M-state. The longitudinal component of $\mathbf{L}$ is hence further reduced. These oscillations cannot be quenched by thermal fluctuations since the energy splitting $E$ is large compared to the thermal energy, even at room temperature. In particular, $E>k_\text{B}T_\text{N}$, where $T_\text{N}$ is the N\'eel temperature. We also note that perturbations $\Delta J$ in classical spin systems at finite temperature give rise only to a rapid relaxation, not coherent oscillations \cite{Hellsvik2016}. Moreover, the oscillations show a macroscopic well-defined phase, since the excitation is impulsive: therefore a macroscopic ensemble measurement can reveal them. In the quantum mechanical scenario we are depicting, the initial phase of the oscillations is determined both by the sign of the $\Delta J$ and the polarization of the optical pump pulse. Although an optical perturbation of the exchange interactions induces in principle a change $\Delta J$ homogenous throughout the system; the pump pulse perturbs the exchange bonds along different crystallographic directions in a non-equivalent way, depending on the direction of the electric field of light (i.e. polarization). The light-matter interaction can also depend on the orientation of the electrical field with respect to the equilibrium orientation of the N\'eel vector. Thus, oscillations induced by pump pulses parallel or perpendicular to the N\'eel vector can have different phases, although in general it is expected that the contribution of the perturbation of exchange interaction dominates. This is because spin-orbit interactions are much weaker than the exchange interactions and perturbation of the spin-orbit coupling alone cannot trigger the purely longitudinal oscillation of the N\'eel vector.}

%\\\\\\\\\\\\\\\\\\\\\\\\\\\\\\\\\\\\\\\\\\\\\\\\\\\\\\\\\\\\\\\\\\\\\\\\\\\\\\\\\\\\\
% Microscopic theory of two-magnon dynamics
%\\\\\\\\\\\\\\\\\\\\\\\\\\\\\\\\\\\\\\\\\\\\\\\\\\\\\\\\\\\\\\\\\\\\\\\\\\\\\\\\\\\\\
%
\subsection{Microscopic theory of two-magnon dynamics}\label{subsection:microscopictheory}
{In this subsection we present a more mathematical treatment of the microscopic quantum description of impulsively stimulated 2M dynamics, similarly to what has been already introduced in the literature \cite{Zhao2004,Zhao2006,Bossini2016} and as outlined qualitatively in the previous subsection. We give a self-contained derivation starting with the perturbation of exchange interactions as the excitation mechanism. Subsequently and beyond the existing theory, we show how the theoretical solution can be simplified by the introduction of Bose pair operators which allow us to link the dynamics of the order parameter with the dynamics of the spin correlations. Such connection was previously shown only at zero temperature.}

\subsubsection{{Effective Hamiltonian of light-matter interaction}}
In the literature, K$_2$NiF$_4$ and KNiF$_3$ are considered as prototype Heisenberg quantum antiferromagnets on simple cubic lattice structures in two and three dimensions, respectively.~\cite{Lines1967} Therefore, we describe the microscopic excitation mechanism of the ISRS on the 2M-mode considering the quantum Heisenberg model with only nearest neighbour exchange interactions parametrised with the constant $J > 0$:

\begin{align}\label{H0}
H_0 = J\sum_{i,\boldsymbol{\delta}}\hat{\mathbf{S}}_i\cdot\hat{\mathbf{S}}_{i+\boldsymbol{\delta}},
\end{align}

\noindent where $\hat{\mathbf{S}}_i$ is the spin operator at site $i$ and $\boldsymbol{\delta}$ is a vector connecting a magnetic site with one of its nearest neighbors on the opposite sublattice. Note that the modulus of $\boldsymbol{\delta}$ is equal to the lattice constant $a$ (see Fig. \ref{fig:CrystalStructure}(a)). { In general, the Raman tensor responsible for 2M scattering can be derived from a symmetry analysis and it represents a light-induced modification of the exchange interaction.~\cite{Fleury1968, Lockwood1987, Zhao2004}}. Here, to facilitate a simple microscopic description of 2M excitation, we discuss light-induced perturbations to the exchange interaction as derived from the electronic Hubbard model.~\cite{Mentink2015,Mikhaylovskiy:2015kz,Itin2016} For a given bond along $\boldsymbol{\delta}$ we have

\begin{align}\label{dHsc}
\Delta J(\boldsymbol{\delta})& =\frac{t^2_0}{2U}\frac{(e\boldsymbol{\delta}\cdot\mathbf{E}_0)^2}{U^2-\hbar^2 \omega^2}.
\end{align}

\noindent where $t_0$ is the hopping integral, $U$ the onsite Coulomb interaction, $e$ the unit charge and $\hbar = h / 2\pi$ with $h$ being the Planck constant. The symbol $\omega$ represents the angular frequency of the electric field of light, while $\boldsymbol{\delta}\cdot\mathbf{E}_0/a$ is the projection of the optical electric field along the nearest-neighbour bond between two spins. This equation reveals how $J$ can be changed by an off-resonant driving of the charge-transfer transition in the Hubbard model. Hence, this approach takes into account the virtual charge-transfer processes between sites belonging to the same band, but not the electric dipole transitions to higher bands. Nevertheless, already from the current model, we observe that the sign of $\Delta J$ is different for off-resonant driving laser pulses with photon-energy tuned below and above the charge-transfer gap. \textcolor{black}{The experiments here reported always employed a driving electric field oscillating at frequencies below the charge-transfer gap and therefore no change of sign of $\Delta J$ is expected from the model.} Extending the model, by including more bands and provided that the symmetry of the crystal allows it, the combination of all transitions can cause the sign of $\Delta J$  to become dependent on the orientation of the electric field of light with respect to the N\'eel vector and the crystallographic axes  (see the phenomenological analysis in Sec.\ref{sub:PumpPolDepTheo}).

Considering exclusively optical perturbations to the exchange, the light matter interaction takes the form

\begin{equation}
\label{dH}
\delta H = \frac{1}{2}f(t)\sum_{i,\boldsymbol{\delta}} \Delta J(\boldsymbol{\delta})\hat{\mathbf{S}_i}\cdot\hat{\mathbf{S}}_{i+\boldsymbol{\delta}},
\end{equation}

\noindent where the function $f(t)$, which is normalized to 1 at its maximum value, describes the time-profile of the light pulse.

\subsubsection{{Magnon modes}}
\textcolor{black}{The strength required to modify $J$ by an amount comparable with the exchange itself is of the same order as the atomic electric field.} As the intensity of the pump signal is smaller than the atomic electric field, we assume that the photoinduced fluctuations are small deviations from the equilibrium state ($\Delta J\ll J$). Such fluctuations can be described in terms of magnon modes. Following the standard approach \cite{fazekas1999}, we introduce bosonic annihilation (creation) operators  $\hat{\alpha}_\mathbf{k}$ ($\hat{\alpha}^\dagger_\mathbf{k}$), $\hat{\beta}_\mathbf{k}$ ($\hat{\beta}^\dagger_\mathbf{k}$), which correspond to two types of magnon modes and  whose detailed derivation is given in Appendix \ref{a:magnonpair}. As aforementioned, the structure of the eigenmodes  strongly depends upon the $\mathbf{k}$-vector.  
In particular, for $\mathbf{k}\approx \mathbf{k}_\mathrm{R}$ operator $\hat{\alpha}^\dagger_\mathbf{k}$ ($\hat{\beta}^\dagger_\mathbf{k}$) creates spin excitations mainly in one magnetic sublattice A (B).

To represent the wavefunction of the magnon modes we use the basis of the Fock states with a fixed number $n_{\uparrow\mathbf{k}}$  and $n_{\downarrow\mathbf{-k}}$ in each mode, where the arrows indicate the two subalattices (to which the magnons belong) and $\mathbf{k}$ is the wavevector. The one-magnon operators act on the Fock states in a standard way:

\begin{equation}
\hat{\alpha}_\mathbf{k}^\dagger|n_{\uparrow\mathbf{k}}\rangle=\sqrt{n_{\uparrow\mathbf{k}}+1}|n_{\uparrow\mathbf{k}}+1\rangle,\quad \hat{\beta}_\mathbf{-k}^\dagger|n_{\downarrow\mathbf{-k}}\rangle=\sqrt{n_{\downarrow\mathbf{-k}}+1}|n_{\downarrow\mathbf{-k}}+1\rangle,
\end{equation}

\noindent and the vacuum states are $\hat{\alpha}_\mathbf{k}|0_{\uparrow\mathbf{k}}\rangle=0$, $\hat{\beta}_\mathbf{-k}|0_{\downarrow\mathbf{-k}}\rangle=0$.

Neglecting magnon-magnon interactions, the Hamiltonians in Eqs. (\ref{H0}) and (\ref{dH}) are expressed as follows:

\begin{eqnarray}\label{eq_AF_hamiltonian_approx}
\hat H_0 &= &E_0 +  \sum_\mathbf{k}\hbar\omega_\mathbf{k}\left[\hat\alpha^\dagger_\mathbf{k}\hat\alpha_\mathbf{k}+\hat\beta^\dagger_\mathbf{k}\hat\beta_\mathbf{k}+1\right], \\ 
\label{eq_AF_hamiltonian_approx2}
\delta\hat H &= &f(t)\sum_\mathbf{k}\left\{\hbar\delta\omega_\mathbf{k}\left[\hat\alpha^\dagger_\mathbf{k}\hat\alpha_\mathbf{k}+\hat\beta^\dagger_\mathbf{k}\hat\beta_\mathbf{k}+1\right]+ \right. \\ \nonumber
&& \qquad\qquad+\left. \hbar V_\mathbf{k}\left[\hat\alpha_{\mathbf{k}}\hat\beta_{-\mathbf{k}} +\hat\alpha^\dagger_{\mathbf{k}}\hat\beta^\dagger_{-\mathbf{k}}\right]\right\},
\end{eqnarray}

\noindent where the constant $E_0=\frac{\hbar}{2}(\Omega + \delta\omega_\mathrm{R})N(S+1)$ sets the reference level energy, $\Omega \equiv z_NJS/\hbar$ ($z_N$ being the number of nearest neighbours) and $\delta\omega_\mathrm{R}\equiv z_N\Delta JS/\hbar$ are the magnon frequency and light-matter coupling constant at the \textit{R}-point, respectively. {We observe that while $\hat{H}_0$ is diagonal in the magnon operators, $\delta{\hat{H}}$ contains also an off-diagonal term.} The term containing $V_\mathbf{k}$ is responsible for the excitation and annihilation of magnon pairs during the action of the pump pulse. The perturbation of the exchange interaction by the optical stimulus induces also an effective shift of the magnon frequency amounting to $\delta \omega_\mathrm{R}$, which is limited to the duration of the pump laser pulse. Hence, the $\delta \omega_\mathrm{R}$ should not be interpreted as a modification of the frequency of the magnons observed in our experimental data at positive delays, but only as an expression in the light-spin interaction. The parameters in Eqs.(\ref{eq_AF_hamiltonian_approx}) and (\ref{eq_AF_hamiltonian_approx2}) are defined as:

\begin{eqnarray}
\label{eq_definition}
\omega_\mathbf{k}&=&\Omega\sqrt{1-\gamma^2_\mathbf{k}},\\
\label{eq_definition2}
\delta\omega_\mathbf{k}&=&\delta\omega_\mathrm{R}\frac{1-\xi_\mathbf{k}\gamma_\mathbf{k}}{\sqrt{1-\gamma^2_\mathbf{k}}}, \\
\label{eq_definition3}
V_\mathbf{k} = V_\mathbf{k}^*&=&\delta\omega_\mathrm{R}\frac{\xi_\mathbf{k}-\gamma_\mathbf{k}}{\sqrt{1-\gamma^2_\mathbf{k}}}.
\end{eqnarray}

\noindent Here the factors $\gamma_\mathbf{k}=\frac{1}{z}\sum_{\boldsymbol{\delta}}\exp(i\mathbf{k}\cdot\mathbf{\boldsymbol{\delta}})$ and $\xi_\mathbf{k}=\frac{1}{za^2}\sum_{\boldsymbol{\delta}}(\hat{e}\cdot\hat{\boldsymbol{\delta}})^2\exp(i\mathbf{k}\cdot\mathbf{\boldsymbol{\delta}})$ depend on the structure of the lattice and on the orientation of the electric field. For a cubic lattice (KNiF$_3$), which is relevant for the experiments here discussed, it follows that:

\begin{equation}\label{eq_gamma_factor}
\gamma_\mathbf{k}=\frac{1}{3}\sum_{j=x,y,z}\cos\left(k_ja\right), \quad
\xi_\mathbf{k}=\frac{1}{3}\sum_{j=x,y,z}e_j^2\cos\left(k_ja\right).
\end{equation}

In a tetragonal (K$_2$NiF$_4$) system we have:

\begin{equation}\label{eq_gamma_factor_K2}
\gamma_\mathbf{k}=\frac{1}{2}\sum_{j=x,y}\cos\left(k_ja\right), \quad
\xi_\mathbf{k}=\frac{1}{2}\sum_{j=x,y}e_j^2\cos\left(k_ja\right).
\end{equation}

%\vspace{0.5cm}
%
%\noindent{\bfseries{Magnon-pair description}}
%\vspace{0.5cm}
\subsection{{2M operators}}
\noindent To solve the spin dynamics triggered by the light-matter interaction described by Eq.~\eqref{eq_AF_hamiltonian_approx2}, it is convenient to introduce operators that directly work on magnon pairs. We define them as

\begin{eqnarray}\label{eq_magnon_pair_operators1}
\hat{K}_\mathbf{k}^z&=&\,(\hat{\alpha}_\mathbf{k}^\dagger\hat{\alpha}_\mathbf{k} + \hat{\beta}_{-\mathbf{k}}^\dagger\hat{\beta}_{-\mathbf{k}}+1)/2,\\\label{eq_magnon_pair_operators2}
\hat{K}_\mathbf{k}^+&=&\,\hat{\alpha}_\mathbf{k}^\dagger\hat{\beta}_{-\mathbf{k}}^\dagger,\quad
\hat{K}_\mathbf{k}^-=\,\hat{\alpha}_\mathbf{k}\hat{\beta}_{-\mathbf{k}},
\end{eqnarray}

\noindent where  $z$ is the quantization axis which coincides with the equilibrium orientation of spins. The physical interpretation of these operators is that $\hat{K}_\mathbf{k}^z$ is the number operator in the two-magnon mode basis, while $\hat{K}_\mathbf{k}^+,\hat{K}_\mathbf{k}^-$ describe creation and annihilation of 2M-mode states, respectively. In the context of a coherent state description, such magnon-pair operators are also known as Perelomov operators,\cite{Perelomov1975,Mattis1981} while in the theory of magnetism they are referred to as hyperbolic operators \footnote{To the best of our knowledge, these magnon-pair operators have not been introduced before in the context of antiferromagnetic magnon theory.}. As it follows from the Bose commutator relations $[\alpha_\mathbf{k},\alpha^\dagger_\mathbf{k}]=1$ and $[\beta_\mathbf{k},\beta^\dagger_\mathbf{k}]=1$, the magnon-pair operators exhibit simple commutation relations:

\begin{equation}\label{e:commutators}
[\hat{K}_\mathbf{k}^z,\hat{K}_\mathbf{k}^\pm] = \pm \hat{K}_\mathbf{k}^\pm, \quad [\hat{K}_\mathbf{k}^-,\hat{K}_\mathbf{k}^+] = 2\hat{K}_\mathbf{k}^z.
\end{equation}

{We also note that the magnon-pair operators are distinct from Schwinger bosons. While both deal with introduction of two types of bosons, the bosons introduced by Schwinger are introduced for each single spin or one single mode, with an additional constraint to satisfy spin conservation, \textit{i.e.} SU(2) symmetry. Instead, the magnon-pair operators combine two bosons that correspond to two different magnon modes.} The role of {spin conservation} is played by the so-called Casimir invariant

\begin{equation}
\label{eq_Casimir}
\hat{Q}_\mathbf{k}=\frac{1}{2}\left( \hat{K}_\mathbf{k}^+\hat{K}_\mathbf{k}^- + \hat{K}_\mathbf{k}^-\hat{K}_\mathbf{k}^+\right)-\left(\hat{K}_\mathbf{k}^z\right)^2 = \frac{1}{4}\left[1-\left(\hat{\alpha}_\mathbf{k}^\dagger\hat{\alpha}_\mathbf{k}-\hat{\beta}_{-\mathbf{k}}^\dagger\hat{\beta}_{-\mathbf{k}}\right)^2\right],
\end{equation}
\vspace{0.1cm}

\noindent which commutes with all $\hat K$ operators. In particular, the conservation law $\Delta S=0$ is respected by the magnon pair operators, since although the total number of magnons $n_{\uparrow\mathbf{k}}+n_{\downarrow\mathbf{-k}}$ can be changed, terms with $\hat{K}_\mathbf{k}^\pm$ change the number of magnons in each of the sublattices by the same amount ($n_{\uparrow\mathbf{k}}=n_{\downarrow\mathbf{-k}}$). In addition, only pairs of magnons of opposite $\mathbf{k}$ are excited, respecting $\Delta\mathbf{k}=0$. In the 2M basis we can express the Casimir invariant as $\hat Q_\mathbf{k}=[1- (n_{\uparrow\mathbf{k}}-n_{\downarrow\mathbf{-k}})^2]/4$. {Mathematically, this conservation law is reflected by the fact that the} commutation relations differ from spin commutation relations and generate the Lie algebra of SU(1,1) instead of SU(2).~\cite{Perelomov1975} In other words, while spin operators describe rotations {constrained to the unit sphere}, {the} operators $\hat K$ describe abstract rotations {constrained to the hyperbolic unit sphere.} {As we will see, physically this corresponds to longitudinal oscillations that conserve the total spin instead of precessions that keep the magnitude of the spins conserved.}

{As a result, to work with} $\hat K$ operators we {can} choose a subspace of the Hilbert space spanned by the vectors $|n_{\uparrow\mathbf{k}}\rangle|n_{\downarrow\mathbf{-k}}\rangle=|n_{\mathbf{k}}\rangle|n_{\mathbf{k}}\rangle$. These 2M states correspond to the discrete representation of the group SU(1,1) for which the Casimir invariant $\hat{Q}=1/4$. The operators $\hat{K}^+_\mathbf{k}$ ($\hat{K}^-_\mathbf{k}$) create (annihilate) the two-magnon states: 

\begin{eqnarray}
\hat K^+_\mathbf{k}|n_{\mathbf{k}}\rangle|n_\mathbf{k}\rangle&=&(n_\mathbf{k}+1)|n_{\mathbf{k}}+1\rangle|n_{\mathbf{k}}+1\rangle,\\ \nonumber
\hat K^-_\mathbf{k}|n_{\mathbf{k}}\rangle|n_\mathbf{k}\rangle&=&n_\mathbf{k}|n_{\mathbf{k}}-1\rangle|n_{\mathbf{k}}-1\rangle,
\end{eqnarray}

\noindent and these states are the eigenstates of the operator $\hat{K}^z_\mathbf{k}$:

\begin{equation}
\hat K^z_\mathbf{k}|n_{\mathbf{k}}\rangle|n_{\mathbf{k}}\rangle=(n_\mathbf{k}+1/2)|n_{\mathbf{k}}\rangle|n_{\mathbf{k}}\rangle.
\end{equation}

\noindent The vacuum states for the magnon-pair operators are defined as follows:

\begin{equation}
\hat K^-_\mathbf{k}|0_{\uparrow\mathbf{k}}\rangle|0_{\downarrow\mathbf{-k}}\rangle=0.
\end{equation}

\noindent{Substituting Eqs.~\eqref{eq_magnon_pair_operators1} and \eqref{eq_magnon_pair_operators2} in the Hamiltonians Eqs. (\ref{H0}) and (\ref{dH}) we obtain:}

\begin{eqnarray}\label{e:H}
\hat H_0 &= &E_0 +  \sum_\mathbf{k}2\hbar\omega_\mathbf{k}\hat{K}_\mathbf{k}^z, \\ 
\label{e:H2}
\delta\hat H &= &f(t)\sum_\mathbf{k}\left[2\hbar\delta\omega_\mathbf{k}\hat{K}_\mathbf{k}^z+ \hbar V_\mathbf{k}\left(\hat{K}_\mathbf{k}^++\hat{K}_\mathbf{k}^-\right)\right].
\end{eqnarray}
In the next section we use the effective Hamiltonians \eqref{e:H} and \eqref{e:H2} for the analysis of light-induced spin dynamics.

\subsubsection{{Spin-spin correlations and the N\'eel vector}}
{The main difference between classical and quantum dynamics arises due to the non-local quantum spin-spin correlations that are neglected in the conventional classical description. A direct link between the longitudinal component of the N\'eel vector and the longitudinal correlators can be obtained within harmonic magnon theory. From the definition Eq. ~(\ref{eq:DefIntro}) we obtain (see Appendix~\ref{a:neelcorr})
	\begin{equation}\label{eq_L_component}
	\hat L_z= \frac{NS}{2} - \frac{1}{z_NS} \sum_{j,\boldsymbol{\delta}} \hat{S}_j^z \hat{S}_{j+\boldsymbol{\delta}}^z,
	\end{equation} 
}{in accordance with what was previously derived~\cite{Bossini2016}. }

{The individual components of the spin correlations can be then expressed in terms of the magnon-pair operators. As detailed in Appendix~\ref{a:magnonpair}, we obtain for the longitudinal spin correlator along the quantization axis the following result}
\begin{eqnarray}\label{eq_longitudinal_correlation}
&&\sum_{j, \boldsymbol{\delta}} \hat{S}_j^z \hat{S}_{j+\boldsymbol{\delta}}^z = -\frac{1}{2}{z_NN}S^2
+z_NS \sum_\mathbf{k} \left(\frac{2\hat{K}_\mathbf{k}^z}{\sqrt{1-\gamma_\mathbf{k}^2}}  -1\right)\nonumber + \\
&&\qquad\qquad\qquad-z_NS \sum_\mathbf{k}\frac{\gamma_\mathbf{k}}{\sqrt{1-\gamma_\mathbf{k}^2}}\left[\hat{K}_\mathbf{k}^++\hat{K}_\mathbf{k}^-\right].
\end{eqnarray}
{Eq.~(\ref{eq_longitudinal_correlation}) shows that the longitudinal spin correlations and hence, $\hat L_z$ includes operators $\hat K_\mathbf{k}^\pm$ which do not commute with $\hat H_0$. In other words, both longitudinal correlators and the N\'eel vector are not conserved quantities and hence can show dynamics even if the total energy is conserved.}

{Instead, the} longitudinal component of the magnetization, like the $\hat Q_\mathbf{k}$ operator, depends on the difference between the number of $\uparrow$ and $\downarrow$ magnons,

\begin{equation}
\hat M_z=\sum_\mathbf{k}(\hat{\alpha}_\mathbf{k}^\dagger\hat{\alpha}_\mathbf{k}-\hat{\beta}_{-\mathbf{k}}^\dagger\hat{\beta}_{-\mathbf{k}}).
\end{equation}

\noindent {We thus find that $[M_z,\hat{Q}_\mathbf{k}]=0$ and in} the case when only 2M Raman processes are considered, $M_z=0$ during the whole dynamics.

It should be noted that the transverse spin components, which would give rise to a magnetization $M_{x,y}$ detectable via Faraday rotation, are expressed through linear combinations of one-magnon mode operators $\hat{\alpha}_\mathbf{k}$, $\hat{\beta}_\mathbf{k}$. {Here and below we} restrict our theoretical model to 2M-Raman processes, $L_{x,y}=M_{x,y}=0$, in accordance with the experimentally demonstrated absence of the Faraday rotation.~\cite{Bossini2016}

\subsubsection{{Impulsively stimulated spin dynamics}}
{The introduced magnon-pair operators allow to model properly the spin dynamics generated by the 2M-mode. In particular, we show that we can describe the dynamics without knowledge of the full initial state, which makes our results applicable both at zero temperature and at finite temperature.} To this end, it is convenient to employ the interaction picture and to introduce the time-dependent operators $\hat K(t)=\exp(i\hat H_0t/\hbar)\hat K\exp(-i\hat H_0t/\hbar)$. Using the commutation relations Eq.~\eqref{e:commutators} we obtain

\begin{equation}\label{eq_K_time_dependent}
\hat{K}_\mathbf{k}^\pm(t)=\hat{K}_\mathbf{k}^\pm e^{\pm \text{i}2\omega_\mathbf{k}t},\quad \hat{K}_\mathbf{k}^z(t)=\hat{K}_\mathbf{k}^z.
\end{equation}

\noindent Note that the operator $\hat{K}_\mathbf{k}^z(t)$, which defines the number of magnons, commutes with $\hat H_0$ and hence its expectation value is time-independent. The time-dependence of a quantum mechanical state is then defined by the unitary evolution operator $\hat U(t,0)$  which satisfies the equation 

\begin{equation}\label{eq_evolution_operator}
i\hbar\frac{d\hat U(t,0)}{dt}=f(t)\sum_\mathbf{k}\left[2\hbar\delta\omega_\mathbf{k}\hat{K}_\mathbf{k}^z+ \hbar V_\mathbf{k}\left(\hat{K}_\mathbf{k}^++\hat{K}_\mathbf{k}^-\right)\right]\hat U(t,0).
\end{equation}

\noindent As in the experiments the duration of the laser pulses $\tau_\mathrm{pulse}$ is considerably smaller than the oscillation period of the magnons: $\tau_\mathrm{pulse} \Omega \ll 1$. So we model the temporal profile of the pump pulses as $f(t)=\tau_\mathrm{pulse}\delta(t)$, where $\delta$ represents the Dirac delta-function. 
{Exploiting that in the harmonic approximation the Hamiltonian is diagonal in $\mathbf{k}$}, the time-evolution operator can be calculated explicitly as follows:

\begin{eqnarray}\label{eq_evolution_operator_delta}
&\hat U(t,0)&=\prod_\mathbf{k}\hat U_\mathbf{k}(t,0),\\
\nonumber&\hat U_\mathbf{k}(t,0)&=\exp\{i\tau_\mathrm{pulse}\left[{V}_\mathbf{k}\left(\hat{K}_\mathbf{k}^+(t)+\hat{K}_\mathbf{k}^-(t)\right)+\delta{\omega}_\mathbf{k}\hat K^z_\mathbf{k}(t)\right]\}.
\end{eqnarray}

\noindent The temporal evolution of the expectation value of  an operator $\hat K_\mathbf{k}$ is then calculated as $K_\mathbf{k}(t)= \langle \hat U^\dag_\mathbf{k}(t,0)\hat K_\mathbf{k}\hat U_\mathbf{k}(t,0)\rangle${, where the symbol $\langle\ldots\rangle$ means quantum-mechanical averaging over the initial state}. Using the commutation relations in Eq.~\eqref{e:commutators} we obtain the  following expressions for the observables $K_\mathbf{k}(t)$:

\begin{eqnarray}\label{eq-observables_average1}
\nonumber K^+_\mathbf{k}(t)+K^-_\mathbf{k}(t)&=&-4{V}_\mathbf{k}\tau_\mathrm{pulse}\langle\hat K^z_\mathbf{k}\rangle\sin(2\omega_\mathbf{k}t-\varphi_\mathbf{k})+\\ 
&&+\langle\hat K^+_\mathbf{k}\rangle e^{2i\varphi_\mathbf{k}}+ \langle\hat K^-_\mathbf{k}\rangle e^{-2i\varphi_\mathbf{k}},\\
K^z_\mathbf{k}(t)&=&\langle\hat K^z_\mathbf{k}\rangle-2{V}_\mathbf{k}\tau_\mathrm{pulse}\langle\hat K^+_\mathbf{k}+\hat K^-_\mathbf{k}\rangle \sin(2\omega_\mathbf{k}t+\varphi_\mathbf{k})\nonumber\\
&&+2i{V}_\mathbf{k}\tau_\mathrm{pulse}\langle\hat K^+_\mathbf{k}-\hat K^-_\mathbf{k}\rangle \cos(2\omega_\mathbf{k}t+\varphi_\mathbf{k}),\label{eq-observables_average2}
\end{eqnarray}

\noindent where the light-induced phase is 

\begin{equation}\label{key}
\varphi_\mathbf{k}=\delta \omega_\mathrm{R}\tau_\mathrm{pulse}\frac{1-\xi_\mathbf{k}\gamma_\mathbf{k}}{\sqrt{1-\gamma_\mathbf{k}^2}},
\end{equation}

\noindent {To arrive at this results we used that} the light-induced modification of the exchange interaction is small compared to the exchange constant itself, i.e., {$\Delta J\ll J\Rightarrow$} $\delta\omega_\mathrm{R}\ll\Omega$ and $\delta\omega_\mathrm{R}\tau_\mathrm{pulse}\ll1$ and kept only the first nontrivial terms linear in $\delta\omega_\mathrm{R}$. {In equilibrium $V_\mathbf{k}=0$ and hence it follows that $K^z_\mathbf{k}(t)=\langle\hat K^z_\mathbf{k}\rangle$, $\hat{K}_\mathbf{k}^\pm(t)=0$.} Substituting Eqs.~(\ref{eq-observables_average1}) {and (\ref{eq-observables_average2}) into Eq. (\ref{eq_longitudinal_correlation}) and using} Eq. (\ref{eq_L_component}) we can formulate the following expression for the time-dependence of the longitudinal component of the N\'eel vector:

\begin{equation}
\label{eq_longitudinal_oscillaitons2}
\langle L_z(t)\rangle=NS-\Delta L(0)-\delta L(t),
\end{equation}

\noindent where 

\begin{equation}\label{eq_Lfluctuations}
\Delta L(0)= \sum_\mathbf{k} \left(\frac{2\langle \hat{K}_\mathbf{k}^z \rangle}{\sqrt{1-\gamma_\mathbf{k}^2}}-1\right)
\end{equation}
\noindent and 

\begin{equation}\label{eq_Loscillations}
\delta L(t)=
4\tau_\mathrm{pulse} \sum_\mathbf{k}\frac{\gamma_\mathbf{k}{V}_\mathbf{k}}{\sqrt{1-\gamma_\mathbf{k}^2}}\langle\hat K^z_\mathbf{k}\rangle\sin(2\omega_\mathbf{k}t-\varphi_\mathbf{k}).
\end{equation} 

{\noindent A number of remarks are in place here. First, we observe that the static value of the N\'eel vector is reduced as compared to the case of maximally aligned spins in each sublattice. This is directly related to the fact that the ground state is dressed by magnons. As is well-known for the quantum antiferromagnet, this is even true in the ground state at $T=0$, when no thermally-induced magnons are present and $\langle \hat{K}_\mathbf{k}^z \rangle=1/2$, which is a consequence of the fact that $\hat{L}_z$ and $\hat{H}_0$ do not commute.} 
{Second, $\delta L(t)$ depends on $K^+_\mathbf{k}(t)+K^-_\mathbf{k}(t)$ (Eq.~\eqref{eq-observables_average1}) and does not require explicit knowledge of the initial state. In particular, it is sufficient to know only the equilibrium value of $\langle \hat{K}_\mathbf{k}^z \rangle$. Hence, our results} are valid also at elevated temperature as long as the harmonic approximation is sufficiently accurate at the {temperature and }wavelength considered. 

It is also instructive to note that the time-dependence of the correlators in Eqs.(\ref{eq_longitudinal_correlation}), with account of the formulas in Eqs.(\ref{eq-observables_average1}) takes the following form:

\begin{eqnarray}\label{eq_correlations}
&&\sum_{j,\boldsymbol{\delta}}\langle \hat{S}_j^z \hat{S}_{j+\boldsymbol{\delta}}^z \rangle\approx
-\frac{1}{2}z_NNS^2+z_NS \Delta L(0)+z_NS\delta L(t)
\end{eqnarray}

%\\\\\\\\\\\\\\\\\\\\\\\\\\\\\\\\\\\\\\\\\\\\\\\\\\\\\\\\\\\\\\\\\\\\\\\\\\\\\\\\\\\\\
% Effective macroscopic theory of two-magnon dynamics
%\\\\\\\\\\\\\\\\\\\\\\\\\\\\\\\\\\\\\\\\\\\\\\\\\\\\\\\\\\\\\\\\\\\\\\\\\\\\\\\\\\\\\

{\subsection{Effective macroscopic theory of 2M dynamics}}
{In the previous subsection we have introduced magnon-pair operators to facilitate a microscopic description of 2M dynamics in the harmonic approximation. In this subsection we will focus on deriving an effective macroscopic theory of 2M dynamics, which supplements the standard Landau-Lifshitz equations on femtosecond time scales. To this end we employ again the magnon-pair operators. In particular, analogous to what has been widely used for the dynamics of quantum spins, we utilize coherent states for magnon pair operators to derive effective macroscopic equations of motion for the dynamics of the antiferromagnetic vector.}
%% - coherent states and quasi-classical dynamics
%% - explicit restriction delta k region
%% - equation of motion
%% - generalized force

%\noindent {\bfseries{Coherent states and quasiclassical dynamics}}

%\vspace{0.5cm}

To introduce the effective macroscopic variables we note that, according to Eq.(\ref{eq_evolution_operator_delta}) the wavefunction after the photo-excitation can be expressed as 

\begin{equation}\label{eq_interaction_picture_wave_function}
|\Psi_\mathrm{AF}(t)\rangle = \prod_\mathbf{k}\hat U_\mathbf{k}(t,0) |\Psi(0)\rangle,
\end{equation}

\noindent where the wavefunction $|\Psi(0)\rangle$ describes the initial state. If the system was initially in the ground state $|\Psi_\mathrm{AF}(0)\rangle =\prod_{\mathbf{k}}|0_{\uparrow\mathbf{k}}\rangle|0_{\downarrow\mathbf{-k}}\rangle$, then, after the pump pulse the wave function in Eq.(\ref{eq_interaction_picture_wave_function}) can be represented as the direct product  $|\Psi(t)\rangle=\prod_{\mathbf{k}}|\mu_{\mathbf{k}}\rangle$ of the  so-called Perelomov's  coherent states:
\begin{equation}\label{eq_coherent_state_2}
|\mu_{\mathbf{k}}\rangle=\sqrt{1-|\mu_{\mathbf{k}}|^2}\sum_{n=0}^\infty\mu_{\mathbf{k}}^n|n_{\mathbf{k}}\rangle|n_{\mathbf{k}}\rangle,
\end{equation}
where the parameter $\mu_{\mathbf{k}}$ is

\begin{align}\label{eq_parameter_coherent_state_2}
\nonumber\mu_{\mathbf{k}}&=i\tanh \left(\delta \omega_\mathrm{R}\tau_\mathrm{pulse}\frac{\xi_\mathbf{k}-\gamma_\mathbf{k}}{\sqrt{1-\gamma_\mathbf{k}^2}}\right) e^{-2i\omega_\mathbf{k}t+i\varphi_\mathbf{k}}\\
\end{align}

\noindent The values of the observables in state $|\Psi_\mathrm{AF}(t)\rangle$ are then obtained by using the general expressions for the averaged values of $\hat K$-operators~\cite{Novaes2004}

\begin{eqnarray}\label{eq_average_K}
\langle\mu_\mathbf{k}|\hat K_\mathbf{k}^-|\mu_\mathbf{k}\rangle&=&\frac{\mu_\mathbf{k}}{1-|\mu_\mathbf{k}|^2},\quad \langle\mu_\mathbf{k}|\hat K^+_\mathbf{k}|\mu_\mathbf{k}\rangle=\frac{\mu_\mathbf{k}^*}{1-|\mu_\mathbf{k}|^2},\nonumber\\
\langle\mu_\mathbf{k}|\hat K_\mathbf{k}^z|\mu_\mathbf{k}\rangle&=&\frac{1}{2}\frac{1+|\mu_\mathbf{k}|^2}{1-|\mu_\mathbf{k}|^2}.
\end{eqnarray}

\noindent The coherent states in Eq. (\ref{eq_coherent_state_2}), like the coherent states in optics, are the closest quantum states to a classical description of the magnonic field. The set of the coherent parameters $\mu_\mathbf{k}$ can thus be considered as proper variables representing the ultrafast spin dynamics of antiferromagnets at a macroscopic level. The equations of motion for these parameters are obtained in a quasiclassical limit as \cite{Bechler2001}:

\begin{equation}\label{eq_quasiclassical}
\partial_t\mu_\mathbf{k}=\{\mu_\mathbf{k},H_\mathrm{class}(\mu_\mathbf{k}, \mu_\mathbf{k}^*)\},
\end{equation}

\noindent where the Poisson brackets $\{\ldots\}$ means

\begin{equation}\label{eq_Poisson_brackets}
\{A,B\}=\frac{(1-|\mu_\mathbf{k}|^2)^2}{i\hbar }\left(\frac{\partial A}{\partial \mu_\mathbf{k}}\frac{\partial B}{\partial \mu_\mathbf{k}^*}-\frac{\partial A}{\partial \mu_\mathbf{k}^*}\frac{\partial B}{\partial \mu_\mathbf{k}}\right).
\end{equation}

The classical Hamiltonian, $H_\mathrm{class}(\mu_\mathbf{k}, \mu_\mathbf{k}^*)\equiv\prod_\mathbf{k}\langle\mu_\mathbf{k}|(\hat H_0+\delta \hat H)\prod_\mathbf{k}|\mu_\mathbf{k}\rangle$, is calculated by substituting the expressions (\ref{eq_average_K}) into (\ref{e:H}). 
{We obtain}

\begin{eqnarray}\label{eq_classical_Hamiltonian}
\nonumber H_\mathrm{class}(\mu, \mu^*)&=&\langle\mu|\hat H|\mu\rangle=\hbar \sum_{\mathbf{k}}\left(\omega_\mathbf{k}+\delta\omega_\mathbf{k}f(t)\right)\frac{1+|\mu_\mathbf{k}|^2}{1-|\mu_\mathbf{k}|^2}+\\ 
&+& V_\mathbf{k}f(t)\frac{\mu_\mathbf{k}+\mu^*_\mathbf{k}}{1-|\mu_\mathbf{k}|^2}.
\end{eqnarray}

{\noindent Recalling} the Hamiltonian in Eq.(\ref{eq_classical_Hamiltonian}), the quasiclassical equations of motion for the parameters $\mu_\mathbf{k}$ take the following form:

\begin{eqnarray}\label{eq_equation_of_motion}
i\hbar\partial_t\mu_\mathbf{k}&=&2\hbar\left(\omega_\mathbf{k}+\delta\omega_\mathbf{k}f(t)\right)\mu_\mathbf{k}+\hbar V_\mathbf{k}f(t)(1+\mu^2_\mathbf{k}).%\\ \nonumber
%		i\hbar\partial_t\mu^*_\mathbf{k}&=&-2\hbar\left(\omega_\mathbf{k}+\delta\omega_\mathbf{k}f(t)\right)\mu^*_\mathbf{k}+\hbar V_\mathbf{k}f(t)(1+(\mu^*_\mathbf{k})^2).
\end{eqnarray}

{\noindent Using (\ref{eq_average_K}), we can directly relate the longitudinal dynamics of the N\'eel vector with the coherent parameters. Combining Eqs. (\ref{eq_average_K}), (\ref{eq_longitudinal_oscillaitons2}) and (\ref{eq_L_component}) we obtain:}
\begin{equation}\label{eq_relation_L_mu_2}
L_z=L_z(0)- \sum_{\mathbf{k}}\frac{\gamma_\mathbf{k}}{\sqrt{1-\gamma_\mathbf{k}^2}}\frac{2\mathrm{Re}\mu_{\mathbf{k}}}{1-|\mu_{\mathbf{k}}|^2}.
\end{equation}

\noindent {Purely longitudinal dynamics of the N\'eel vector, \textit{i.e.} dynamics that does not induce any change of the total magnetization cannot be described by the standard Landau-Lifshitz equations. Therefore, additional macroscopic variables are required. To this end, we propose to use the parameters $\mu_\mathbf{k}$ to characterize short-range spin correlations and femtosecond scale dynamics.}

Equations (\ref{eq_equation_of_motion}) together with the relations (\ref{eq_relation_L_mu_2}) and the standard Landau-Lifshitz equations for magnetic sublattices form a closed set of dynamic equations for macroscopic variables. {Note that, however,} quantum and classical spin dynamics can be disentangled, since they occur at different time-scales. {For example, in KNiF$_3$}, the longitudinal oscillations of $L_z$ take place on a sub-picosecond timescale when the orientation of the N\'eel vector can be considered static. {On the other hand, in the same material t}he classical precessional spin dynamics can be observed on a characteristic time-scale of 100 ps.~\cite{Bossini2014}

{Furthermore, using the quasi-classical equation of motion, Eq.~(\ref{eq_equation_of_motion}) and the Poisson brackets Eq.~\eqref{eq_Poisson_brackets} we can define a generalized force as the conjugate variable of the coherent state variable $\mu_\mathbf{k}$:
	\begin{equation}\label{eq_eff_field1}
	H_{\mu_{\mathbf{k}}}={(1-|\mu_\mathbf{k}|^2)^2}\frac{\partial H_\mathrm{class}}{\partial{\mu_\mathbf{k}^*}}.
	\end{equation}
	\noindent The prefactor $(1-|\mu_\mathbf{k}|^2)^2$ arises because of the curvature of the hyper-unit sphere. The generalized force defined here is the mathematical analogue of the effective magnetic field in the Landau-Lifshitz equations. It allows us to write Eq.~\eqref{eq_equation_of_motion} compactly as
	\begin{equation}\label{eq_equation_of_motion2}
	i\hbar\partial_t\mu_\mathbf{k}=H_{\mu_{\mathbf{k}}}.
	\end{equation}
	This representation is very useful since it enables the treatment of 2M dynamics on a purely phenomenological basis, without resorting to a specific microscopic model, as we will exploit in Sec.~\ref{sub:PumpPolDepTheo}.}

{The parameter $\mu_\mathbf{k}$ in Eq.~(\ref{eq_parameter_coherent_state_2}) is similar to the parameter of the coherent state (states corresponding to minimal uncertainity) often used in quantum optics.  Its modulus is related to the average number of magnon pairs:
	\begin{equation}\label{eq_average-number}
	\overline{n_\mathbf{k}}=\frac{|\mu_{\mathbf{k}}|^2}{1-|\mu_{\mathbf{k}}|^2},
	\end{equation}
	and to the probability to observe $n_\mathbf{k}$ magnon state in each of the correlated magnon modes:
	\begin{equation}\label{eq_probability}
	Prob(n_\mathbf{k})=(1-|\mu_{\mathbf{k}}|^2)|\mu_{\mathbf{k}}|^{2n_\mathbf{k}}=\frac{\overline{n_\mathbf{k}}^{n_\mathbf{k}}}{(1+\overline{n_\mathbf{k}})^{n_\mathbf{k}+1}}.
	\end{equation}
	The values of $\mu_\mathbf{k}$ are related directly with the amplitude and phase  of the oscillation of the N\'eel vector, as it can be seen from the expression (\ref{eq_relation_L_mu_2}).}

{To conclude this subsection, we mention that the q}uasiclassical Eq.~(\ref{eq_equation_of_motion}), or equivalently Eq.~\eqref{eq_equation_of_motion2}, allows to take into account the main features of the {sub-picosecond} description of the antiferromagnetic dynamics. {For example,} it can reproduce the interference between magnonic oscillations induced by two delayed pump pulses, experimentally observed in Ref. \cite{Bossini2016}. To illustrate this aspect, we consider the quantum dynamics induced by two subsequent pulses delayed in time by $t_\mathrm{delay}$, so that $f(t)=\delta(t)+\delta(t- t_\mathrm{delay})$. In the initial state $\mu_\mathbf{k}=0$. As it follows from Eq.~(\ref{eq_equation_of_motion}), after the second pulse 

\begin{equation}\label{eq_solution_mu_two_pulses}
\mu_\mathbf{k}(t)=2 i\tanh V_\mathbf{k} \cos^2(\omega_\mathbf{k} t_\mathrm{delay})e^{-2i\omega_\mathbf{k}t},\quad t\ge t_\mathrm{delay},
%\mu_\mathbf{k}(t)=i\tanh V_\mathbf{k} \cos^2(\omega_\mathbf{k} t_\mathrm{delay})\exp^{\left(-2i\omega_\mathbf{k}t-iV_\mathbf{k}\coth{(2V_\mathbf{k}\cos^2(\omega_\mathbf{k} t_\mathrm{delay}))\sin 2\omega_\mathbf{k}\Delta t_\mathrm{delay}\right)}
\end{equation}

\noindent and, correspondingly, the oscillation amplitude in Eq. (\ref{eq_L_z_approximate}) acquires an additional factor which depends on the time delay in the following way:

\begin{equation}\label{eq_solution_L_two_pulses}
\delta L \Rightarrow  2 \delta L\cos^2(\Omega t_\mathrm{delay}).
\end{equation}

\noindent Equation (\ref{eq_solution_L_two_pulses}) shows that depending on $t_\mathrm{delay}$ the amplitude of the quantum oscillations may be doubled (when $ t_\mathrm{delay}=\pi/\Omega$) or vanish (when $ t_\mathrm{delay}=\pi/2\Omega$) as a result of constructive/destructive interference.

%\\\\\\\\\\\\\\\\\\\\\\\\\\\\\\\\\\\\\\\\\\\\\\\\\\\\\\\\\\\\\\\\\\\\\\\\\\\\\\\\\\\\\
% Pump polarization dependence
%\\\\\\\\\\\\\\\\\\\\\\\\\\\\\\\\\\\\\\\\\\\\\\\\\\\\\\\\\\\\\\\\\\\\\\\\\\\\\\\\\\\\\
\vspace{0.5cm}
\subsection{Pump polarization dependence}
%% - effect sign of Delta J
%% - symmetry argument BZ
%% - generalized light-matter interaction based on symmetry relations

\label{sub:PumpPolDepTheo}
Earlier experiments demonstrated that the initial phase of the oscillations of the antiferromagnetic vector triggered by photo-excitation of the 2M-mode depends on the polarization of the pump.~\cite{Bossini2016} In this Section we try to reveal the origin of such a dependence, based on our microscopic formalism and on phenomenology. 

It follows from our microscopic model (see Eq. (\ref{dHsc}) {and the discussion in Sec.~\ref{subsecexcimech}}) that for a given orientation of the electric field of light $\mathbf{E}$, the initial phase of the oscillations of $\mathbf{L}$ is directly related to the sign of $\Delta J$: an enhancement and a reduction of $J$ generate oscillations of the order parameter with opposite sign. In fact, the term containing ${V}_\mathbf{k}$ in Eqs. (\ref{eq_AF_hamiltonian_approx2}) and (\ref{e:H2}), which is responsible for the excitation of magnon pairs, depends linearly on $\Delta J$. Moreover Eq. (\ref{eq_Loscillations}) shows that $\delta L(t) \propto {V}_\mathbf{k}$ and hence a change in the sign of $\Delta J$ induces a modification of the sign of the amplitude of the antiferromagnetic vector. In our experiment the sign of $\Delta J$ is positive and constant, but it could be negative if a pump photon energy bigger than the band-gap was employed (i.e. $\hbar \omega > U$, see Eq. (\ref{dHsc})).
In addition, Eq. (\ref{eq_Loscillations}) contains a phase factor $\varphi_k$ defined in terms $\delta\omega_R$ (see Eq. (\ref{key})), which depends on $\Delta J$ as well. For a simple cubic lattice Eq. (\ref{dHsc}) shows that the sign of $\Delta J$ is unaffected by rotating the polarization between directions parallel to different crystallographic axes. Thus within this approximation the phase of the signal is independent of the light polarization. 
However, a magnetoelastic strain \cite{Money1980,Ganot1982} or the asymmetry of the electronic orbitals due to spin-orbit interactions can further split degeneracy of $\Delta J$  depending on the orientation of the electric field of the pump with respect to the equilibrium orientation of spins. As a result, the effect of light on spins should be different in the cases when the pump polarization is parallel or perpendicular to the quantization axis (equilibrium orientation of spins). In such a situation, symmetry allows the phase to be different for different orientations of the electric field, because the modification of the exchange interaction induced by light with the electric field parallel to the equilibrium orientation of spins $\Delta J_{||}$ differs from the effect obtained if the polarization of light is rotated by 90 degrees with respect to the spin direction, i.e. $\Delta J_{||}\neq\Delta J_\perp$.

Our description of the light-induced spin dynamics has so far neglected any anisotropy (i.e. $\Delta J_{||}=\Delta J_\perp$).  Considering now this anisotropic contribution, the amplitude and phase of the oscillations can be written as  

\begin{align}\label{eq_polarization}
\nonumber\Delta L&=\frac{2}{3\pi^3}\left({\delta \omega_\mathrm{R}^\perp e^2_\perp+\delta \omega_\mathrm{R}^\|e^2_\|}\right)(\Delta ka)^3\tau_\mathrm{pulse}, \\ 
\varphi&\approx \left({\delta \omega_\mathrm{R}^\perp e^2_\perp+\delta \omega_\mathrm{R}^\| e^2_\|}\right)\tau_\mathrm{pulse},
\end{align}

\noindent \textcolor{black}{where $\delta \omega_\mathrm{R}^\perp$ and $\delta \omega_\mathrm{R}^\|$ represent the components of the modification of the light-matter interaction along a direction perpendicular and parallel to the equilibrium orientation of spins, respectively.} Here the components $e_\perp$ and $e_\|$  of the unit vector of light polarization are projected to the directions perpendicular and parallel to the N\'eel vector.

It is clear from Eq.~(\ref{eq_polarization}) that controlling the light polarization allows one to manipulate the phase and amplitude of the oscillations. The variation of the phase is much more pronounced than the modification of the amplitude. The maximal variation of phase and amplitude $\propto (\delta \omega_\mathrm{R}^\|-\delta \omega_\mathrm{R}^\perp)$ can be achieved by rotating the light polarization from the parallel  ($e_\|=1$, $e_\perp=0$) to the perpendicular  ($e_\|=0$, $e_\perp=1$) configuration. Any rotation of the light polarization which preserves the relation between $e_\|^2$ and $e_\perp^2$ (e.g., within the plane perpendicular to the N\'eel vector) has no effect on the phase of the N\'eel vector oscillations.    

A more general way to describe the effects of the polarization dependence of the longitudinal oscillations of $\mathbf{L}$ is based on a phenomenological modelling of the light-matter interaction. This approach does not specify microscopic mechanisms and takes into account just the symmetry of the sample.~\cite{Zhao2004,Zhao2006} The main idea consists in showing that the observed polarization dependence of the photo-induced change of the correlation function $\langle S^\Uparrow S^\Downarrow\rangle$ is allowed by the symmetry of the crystal. In particular, the 2M-process in antiferromagnets can be described by means of the following phenomenological potential

\begin{equation}\label{eq_potential}
\Phi=\chi_{jklm}E_jE_k\langle S^\Uparrow_l S^\Downarrow_m\rangle,
\end{equation}

\noindent where $\chi_{jklm}$ represents a fourth rank magneto-optical polar tensor and plays the role of magneto-optical susceptibility. $E_{j,k}$ are the amplitude components of the electric field of the pump beam and $\langle S^\Uparrow_l S^\Downarrow_m\rangle$  is the correlation function between spins belonging to different sublattices.

The tensor $\hat \chi$ reflects the symmetry of light-matter interaction and is obviously invariant under permutation of the first two indices, $\chi_{jklm}=\chi_{kjlm}$. The permutation of the second pair of indices is related to the permutation of the magnetic sublattices and thus should be treated according to the magnetic symmetry of the system. Note that Eq. (\ref{eq_potential}) is also a phenomenological description valid for photo-induced magnetic order in paramagnetic media.  To understand how light acts on a magnetically ordered medium, it is important to observe that the structure of the $\chi_{jklm}$ tensor is governed by the symmetry of the \textit{magnetically ordered} phase, which is lower than the symmetry of the paramagnetic phase. For the case, of KNiF$_3$ studied in Ref. \cite{Bossini2016}, the crystallographic point group is $m3m$, however the magnetic order lowers the symmetry of the medium down to $4/mmm$, where the 4-fold axis is along the antiferromagnetic vector. Alternatively, the effect of light on the spin-correlation function in a magnetically ordered medium can be described as a higher order effect: 

\begin{equation}
\Phi = \chi_{jklmno}E_{j}E_{k} \langle S^\Uparrow_l S^\Downarrow_m \rangle L_{n}L_{o},
\end{equation}

\noindent where $\chi_{jklmno}$ is a sixth rank tensor for the $m3m$ crystallographic point group and the form of tensor $\chi_{jklm}$ for the $4/mmm$ point group can be found as $\chi_{jklm} = \chi_{jklmno}L_{n}L_{o}$.

Analyzing the relations between the tensor components of the 4$/mmm$  point group,~\cite{Birss1966} we note the following non-vanishing components of $\hat \chi$ in Voigt notations\footnote{We use Voigt notation for pair indices: $xx \rightarrow 1$, $yy \rightarrow 2$, $zz \rightarrow 3$, $yz \rightarrow 4$, $zx \rightarrow 5$, $xy \rightarrow 6$.}:  $\chi_{11}=\chi_{22}=\chi_{12}=\chi_{21}$, $\chi_{33}$,  $\chi_{13}=\chi_{23}$, $\chi_{31}=\chi_{32}$, $\chi_{44}=\chi_{55}$, and $\chi_{66}$. This means that the function in Eq. (\ref{eq_potential}) can be written as

\begin{eqnarray}\label{eq_potential_tetragonal}
\Phi&=&\chi_{11}\left(E^2_x+E^2_y\right)\left(\langle S^\Uparrow_x S^\Downarrow_x\rangle+\langle S^\Uparrow_y S^\Downarrow_y\rangle\right)+\chi_{33}E^2_z\langle S^\Uparrow_z S^\Downarrow_z\rangle\nonumber\\
&+&\chi_{13}\left(E^2_x+E^2_y\right)\langle S^\Uparrow_z S^\Downarrow_z\rangle+\chi_{31}E^2_z\left(\langle S^\Uparrow_x S^\Downarrow_x\rangle+\langle S^\Uparrow_y S^\Downarrow_y\rangle\right)\nonumber\\
&+&\chi_{66}E_xE_y\langle S^\Uparrow_x S^\Downarrow_y\rangle+\chi_{44}E_z\left(E_x\langle S^\Uparrow_x S^\Downarrow_z\rangle+E_y\langle S^\Uparrow_y S^\Downarrow_z\rangle\right)
\end{eqnarray}
\vspace{0.15cm}	

\noindent where we assumed the additional symmetry of the correlations $\langle S^\Uparrow_l S^\Downarrow_m\rangle=\langle S^\Uparrow_m S^\Downarrow_l\rangle$. The analysis of the function in Eq.(\ref{eq_potential}) for a given polarization state can reveal which correlations could be excited. For this purpose we need to express $\langle S^\Uparrow_l S^\Downarrow_m\rangle$ in terms of parameters of coherent states $\mu_\mathbf{k}$. As the density of light-induced magnonic states is peaked near $\mathbf{k}\approx \mathbf{k}_\mathrm{R}$, we can limit our description to $\mu_{ \mathbf{k}_\mathrm{R}}\equiv \mu_R$. {Moreover, as the pump pulse does not generate one-magnon excitations, all nondiagonal correlations  $\langle S^\Uparrow_x S^\Downarrow_y\rangle$, $\langle S^\Uparrow_x S^\Downarrow_z\rangle$ etc vanish.} Taking into account that the light-induced contribution to the correlation functions can be written as (introducing the phenomenological constant $A$)

\begin{eqnarray}\label{eq_correlation_functions}
\langle S^\Uparrow_z S^\Downarrow_z\rangle&=&A\frac{1+|\mu_R|^2}{1-|\mu_R|^2}+\frac{\mu_R+\mu^*_R}{1-|\mu_R|^2},\\
\langle S^\Uparrow_x S^\Downarrow_x\rangle&=&\langle S^\Uparrow_y S^\Downarrow_y\rangle =\frac{1}{2}\left[(1-A)\frac{1+|\mu_R|^2}{1-|\mu_R|^2}-\frac{\mu_R+\mu^*_R}{1-|\mu_R|^2}\right]%\nonumber
\end{eqnarray}

\noindent we get for the function in Eq.(\ref{eq_potential_tetragonal}) the following phenomenological expression:

\begin{eqnarray}
&\Phi=\frac{\mu_R+\mu^*_R}{1-|\mu_R|^2}E^2\left[\left(\chi_{13}-\chi_{11}\right)e^2_\perp+\left(\chi_{33}-\chi_{31}\right)e^2_\|\right] \\
&+\frac{1+|\mu_R|^2}{1-|\mu_R|^2}E^2\left\{\left[\chi_{13}A+\chi_{11}(1-A)\right]e^2_\perp+ \right.\\ 
&+\left. \left(\chi_{33}A+\chi_{31}(1-A)\right)e^2_\|\right\}.
\label{eq_potential_tetragonal_final}
\end{eqnarray}

\noindent {Using this phenomenological potential we can exploit Eq.~(\ref{eq_eff_field1}) to evaluate the generalized force.} At the leading order in the small parameter $\mu$ we obtain

\begin{equation}\label{eq_effective field}
H_\mu \approx \frac{\partial \Phi}{\partial \mu_R}\approx E^2\left[\left(\chi_{13}-\chi_{11}\right)e^2_\perp+\left(\chi_{33}-\chi_{31}\right)e^2_\|\right].
\end{equation}

\noindent For a short pulse the value of the generalized force defines the initial amplitude and phase of $\Delta L$. From Eq.~(\ref{eq_effective field}) it follows that the symmetry allows different values of $H_{\mu}$ for the parallel ($e_\|=1$, $e_\perp=0$)  and perpendicular ($e_\|=0$, $e_\perp=1$)  configurations. Considering Eq.~(\ref{eq_effective field}) we can formulate the following predictions:

\begin{itemize}
	\item if light propagates along the N\'eel vector, as in the case of KNi$_2$F$_4$, $e_\|=0$, $\Delta L\propto H_\mu\propto \left(\chi_{13}-\chi_{11}\right)$ and does not depend upon the direction of light polarization. 
	\item If light propagates perpendicularly to the N\'eel vector, as in the case of KNiF$_3$, both $e_\|$ and $e_\perp$ components of the polarization vector could be nonzero. The difference between two orthogonal polarization states $\mathbf{E}\|(e_\|, e_\perp)$ and $\mathbf{E}\|(- e_\perp,e_\|)$ is $\propto (\chi_{33}-\chi_{31}-\chi_{13}+\chi_{11})(e^2_\|-e^2_\perp)$. It is maximal when light is polarized parallel/perpendicular to the N\'eel vector.
\end{itemize}

%\\\\\\\\\\\\\\\\\\\\\\\\\\\\\\\\\\\\\\\\\\\\\\\\\\\\\\\\\\\\\\\\\\\\\\\\\\\\\\\\\\\\\
% Quantum aspects of two-magnon dynamics
%\\\\\\\\\\\\\\\\\\\\\\\\\\\\\\\\\\\\\\\\\\\\\\\\\\\\\\\\\\\\\\\\\\\\\\\\\\\\\\\\\\\\\

\subsection{Quantum aspects of 2M dynamics}
%% - squeezing
%% - no fluctuations |m_i^2|
%% - ADD Bell state
%% - ADD entanglement entropy
%% - EXTEND discussion 1/2z
%% - MOVE competition between Ising and spin flip terms absent in classical AFM
%% - MOVE total energy conserved
{In the previous subsections we have introduced qualitative, microscopic and effectively macroscopic descriptions of 2M dynamics, respectively. In this subsection we elaborate on these descriptions focusing on the quantum aspects of 2M dynamics and discuss the relation with squeezing of quantum noise discussed before in connection with 2M excitation. Finally, we elaborate on the role of quantum and thermal fluctuations.} 

% Classical spinwaves: anisotropy and exchange
% Coherent two-magnon dynamics: only exchange - what are the competing interactions?
{A simple perspective on the quantum nature of 2M dynamics follows from analyzing the microscopic interactions involved. For homogenous spin precession in antiferromagnets, the classical calculation gives exactly the same resonance frequencies as the fully microscopic quantum derivation. In both cases the frequency is determined from the competition of anisotropy and inter-sublattice exchange interactions. For 2M dynamics the situation is different. As discussed in Sec.~\ref{subsecexcimech} and from the microscopic theory outlined above, the homogenous dynamics of the N\'eel vector arises from the time-dependent} {quantum oscillations between the states with different numbers of 2M excitations and, correspondingly, with different energies. Such oscillations are similar to Rabi oscillations.}

{Moreover, the quantum states (\ref{eq_coherent_state_2}) which govern dynamics of the N\'eel vector, have all the features of the entangled (non-local and nonseparable) quantum states. First, the state of the system is formed by the pairs of magnons belonging to different modes (spin-up and spin down) and propagating in opposite directions, as can be seen from the following expression:
	\begin{equation}\label{eq_coherent_state_4}
	|\mu_{\mathbf{k}}\rangle=\sqrt{1-|\mu_{\mathbf{k}}|^2}\sum_{n=0}^\infty\mu_{\mathbf{k}}^n|n_{\uparrow\mathbf{k}}\rangle|n_{\downarrow-\mathbf{k}}\rangle\delta(n_{\uparrow\mathbf{k}}-n_{\downarrow-\mathbf{k}}).
	\end{equation}
	This means that, at least theoretically, the individual magnons from different modes can be detected separately. Thus, for each $\mathbf{k}$ the system can be considered as a bipartite. Second, Eq.~(\ref{eq_coherent_state_2}) predicts correlated statistics of the $|n_{\uparrow\mathbf{k}}\rangle$ and $|n_{\downarrow-\mathbf{k}}\rangle$ states of the individual magnon modes  with  equal $n_{\uparrow\mathbf{k}}=n_{\downarrow-\mathbf{k}}=n_{\uparrow\mathbf{k}}$ (see Eq.~(\ref{eq_probability})). In other words, although individual measurement of one magnon mode can detect the state with any possible $n_{\uparrow\mathbf{k}}$, the outcome of the combined (simultaneous) measurement of two magnon modes  is  limited to the states with the same $n_{\uparrow\mathbf{k}}$. This means that the state of the system cannot be represented as a product of independent pure states of each magnon modes (nonseparability). }

Thus, the observed oscillations between the vacuum states $|0_{\uparrow\mathbf{k}}\rangle|0_{\downarrow\mathbf{-k}}\rangle$ and excited states $|n_{\uparrow\mathbf{k}}\rangle|n_{\downarrow\mathbf{-k}}\rangle$ could be treated as indication of the entanglement between $\uparrow$ and $\downarrow$ magnon modes. These states are equivalent to the two-mode coherently-correlated photon states. \cite{Usenko2007} In analogy with optics, where the entangled states are produced by parametric downconversion, correlations between two different magnons modes are established in the course of a second-order magnetic  Raman process,~\cite{Fleury1968} which conserves the total spin of the system. So, oscillations of the N\'eel vector result from the quantum correlations between the spin states at the different magnetic sublattices and have no counterpart in the magnetic dynamics described by the Landau-Lifshitz equations. {We note further that,} in the particular case of femto-nanomagnons, i.e. magnons with $\mathbf{k}\approx\mathbf{k}_\mathrm{R}$, all the coherent states $|\mu_{\mathbf{k}}\rangle$ have almost the same frequency and phase with difference only in amplitude. This additional, "classical" coherence of the different coherent states results in an ensemble response obtained by the sum of contributions from different modes and it allows a macroscopic observation of the quantum effect via optical methods.

{Next, we elaborate on another quantum aspect of two-magnon dynamics, which is the squeezing of quantum noise. The coherent state of the two-magnon mode has been identified as a squeezed state and the fluctuations of the total squared magnetization have been ascribed as the squeezing variable.~\cite{Zhao2004,Zhao2006}. While we agree that the two-magnon dynamics can be interpreted as a squeezed state, we put forward a different variable as squeezing variable, as we will explain below.}
{First, we note that} there are different definitions of squeezed and coherent states,~\cite{Fu1996} which are equivalent only for a harmonic oscillator described by single-mode bosonic operators. In particular, coherent single-mode photon states are simultaneously eigenstates of the annihilation operator and the minimum uncertainty states. In addition, the squeezing modifies the coherent states in a way which reduces quantum fluctuations of one of non-commuting observables below the uncertainty limit. However, in the case of Perelomov's states, such simple classification can be misleading.  In particular, the coherent two-magnon states in Eq.(\ref{eq_coherent_state_2}) are neither eigenstates of {the} operator $\hat K^-$ nor the minimum uncertainty states for arbitrary $\mu$.~\cite{Buzek1990} Second, usually squeezing is associated with transformation of one time-independent state into another time-independent state with reduced fluctuations, which does not apply to the description of time-resolved experiments where $\mu$ is time-dependent. 

However, within a certain extent we can consider the coherent two-magnon states  in Eq.(\ref{eq_coherent_state_2})  as squeezed, meaning that in these states the quantum fluctuations {$\Delta S$ are reduced with respect to their value in the ground state. In particular, in the ground state we have}: $\langle S_{Ai}^z\rangle= - \langle S_{Bj}^z \rangle = S -\Delta S$, where 

\begin{equation}
\Delta S  \equiv\frac{\Delta L(0)}{N}=\frac{1}{N}\sum_\mathbf{k}\left(\frac{1}{\sqrt{1-\gamma_k^2}}-1\right).
\end{equation}

\noindent{As follows from our theory (see Eqs.(\ref{eq_longitudinal_oscillaitons2}) and (\ref{eq_Loscillations})), the longitudinal oscillations of $L_z$ periodically \textit{reduce} $\Delta S$ below the ground-state value.} Thus {$\Delta S$} is identified as the squeezing variable in this context. \textcolor{black}{The definition of the local magnetization has instead some subtleties. In \cite{Zhao2004,Zhao2006}, the local magnetization operator is defined as $\mathbf{m}_i=g\mu_\mathrm{B}(\mathbf{S}_i^\Uparrow + \mathbf{S}_{i+\boldsymbol{\delta}}^\Downarrow)$. Here $\mathbf{S}_{i+\boldsymbol{\delta}}^\Downarrow$ is the spin operator of a nearest neighbour relative to $\mathbf{S}_{i}^\Uparrow$, in the opposite sublattice. A particular choice for $\boldsymbol\delta$, \textit{e.g.} $\boldsymbol\delta_1=a\mathbf{e}_x$ defines an observable with a lower symmetry than the Hamiltonian. The total magnetization $\mathbf{M}=\sum_i\mathbf{m}_i$ as well as the antiferromagnetic order parameter $\mathbf{L}$ and $\Delta S$ are independent of the choice of $\boldsymbol\delta_1$, since these are global variables. However, different choices for $\boldsymbol\delta_1$ can give different results for the local fluctuations $\langle\mathbf{m}_i^2\rangle$. We argue that this merely reflects the choice of the observable rather than the intrinsic physics of the system. To illustrate this, by averaging over $\boldsymbol\delta$ we obtain}

\textcolor{black}{\begin{align}\label{eq-flucenergy}
\frac{1}{(g\mu_\mathrm{B})^2}\sum_{i,\boldsymbol\delta}\langle \mathbf{m}_i^2\rangle&=\sum_{i,\boldsymbol\delta}\langle(\mathbf{S}_i+\mathbf{S}_{i+\boldsymbol\delta})^2\rangle \nonumber \\
&= \sum_{i,\boldsymbol\delta} \langle\mathbf{S}_i^2\rangle + \langle\mathbf{S}_{i+\boldsymbol\delta}^2\rangle + 2\langle\mathbf{S}_i\cdot\mathbf{S}_{i+\boldsymbol\delta}\rangle \nonumber\\
&=NS(S+1)+ 2\sum_{i,\boldsymbol\delta} \langle\mathbf{S}_i\cdot\mathbf{S}_{i+\boldsymbol\delta}\rangle.
\end{align}}

\noindent \textcolor{black}{The last term on the right hand side is proportional to the total energy, which is conserved after the pulse and hence does not show the characteristic oscillation with the period of the 2M mode. Therefore, $\langle\mathbf{m}_i^2\rangle$ for a given $\boldsymbol\delta_1$ may oscillate, but the terms with different $\boldsymbol\delta_1$ cancel each other. Despite this subtlety due to the choice for $\boldsymbol\delta_1$, we note that the claim of Ref.~\cite{Zhao2006} is that $\langle\mathbf{m}_i^2\rangle$ is proportional to $\frac{1}{2}\sum_{j,\boldsymbol{\delta}} \langle\hat S_j^+\hat S_{j+\boldsymbol{\delta}}^- + \hat{S}_{j+\boldsymbol{\delta}}^- \hat{S}_j^+\rangle$. The time-dependent contribution of this term is proportional to $\hat{K}^+(t)+\hat{K}^-(t)$, in fact it has the opposite sign as compared to the longitudinal correlations due to total energy conservation. Hence, the same term is determining the quantum fluctuations in both interpretations.} {To further support the interpretation of $\Delta S$ as squeezing variable we} recall the discussion in Ref. ~\cite{fazekas1999} on the analogy between quantum noise in antiferromagnets and the influence of zero-point motion on the lattice degrees of freedom. For the latter, the quantum noise is characterized by the deviation from the classical position $\delta a = \sqrt{\langle(R_j - \langle R_j \rangle)^2\rangle}$ , which is removed in the limit of infinite mass ($m \rightarrow \infty$). Similarly, the quantum noise in antiferromagnets is characterized by fluctuations of spin $\Delta S$, and the total spin S plays the role of a mass. So, when $S \rightarrow \infty$, $\Delta S \rightarrow0$ and the spin system can be described with classical spins. {To conclude the discussion of squeezing, }we note that fundamental excitations can exhibit also a different kind of squeezing, completely of thermal origin. Experimental observations of thermal squeezing of lattice modes (but not of magnons) on the picosecond time-scale have been reported.~\cite{Trigo:2013fi,Johnson:2009jx}
\\
{\indent We conclude this subsection with a qualitative discussion on the role of quantum and thermal fluctuations.} The main contribution to the quantum noise in the ground state originates from the long-wavelength magnons with $\mathbf{k}\rightarrow 0$ for which $\gamma_\mathbf{k}\rightarrow 1$. {In addition, q}uantum fluctuations to the sublattice magnetization scales as $1/2z_N$,~\cite{fazekas1999} ($z_N$ being the number of nearest neighbors) making it almost undetectable in {three dimensional} samples as {$\Delta S$ is only a few percent of $S$}. On the other hand, light-induced oscillations of the spin-correlations (see Eq. (\ref{eq_correlations})) are related to femto-nanomagnons with $\mathbf{k}$ close to the edges of the Brillouin zone. The number of magnons is proportional to the intensity of light and can thus be detected in macroscopic samples. In addition, while the observation of the quantum noise effects in the ground state demands special conditions (e.g. low temperature), the light-induced 2M oscillations (see Eq. (\ref{eq_Loscillations})) can be induced even if the system was initially in a mixed state. We would like to underline that our theory requires the presence of local rather than long-range order. Our approach can thus be applied at finite temperature: since it is a model based on spin-wave theory it is applicable in the temperature regime $T<< 4T_N$ in 2D and was found to give rather accurate agreement even for temperatures above $T_N$.~\cite{Fleury1970} 

Although calculating the actual temperature dependence of the spin dynamics induced by the excitation of femto-nanomagnons goes beyond the harmonic approximation considered here, we can gain some insight into the role of temperature by addressing the problem in terms of statistical ensemble. In particular, reminding that the dominant contribution to the 2M-process originates from the regions close to the edges of the Brillouin zone, where the magnon density of states peaks, we can further simplify the formulas for the spin dynamics by restricting our interest to the small range $\Delta k\ll \pi/(2a)$ in the  vicinity of the $R$-point. Here $\gamma_\mathbf{k},\xi_\mathbf{k}\propto (\Delta ka)$ and 

\begin{align}\label{eq_approximate_definitions}
\nonumber \omega_\mathbf{k}&\approx\Omega\left[1-O(\Delta k^2a^2)\right], \,\delta\omega_\mathbf{k}\approx \delta\omega_\mathrm{R}\left[1-O(\Delta k^2a^2)\right],\\
V_\mathbf{k}&\propto\delta\omega_\mathrm{R}(\Delta k a).
\end{align}

\noindent Hence, the relevant energy scale for the longitudinal dynamics is $\hbar\omega_R\sim 2zJS$ defined by the high-wavevector magnons, leading to oscillations on the femtosecond time-scale. This is consistent with both the time-domain observations reported here and previously~\cite{Zhao2004,Zhao2006,Bossini2016,Bossini:2016jmbaca,Bossini:2017bo} and with the extensive spontaneous Raman literature~\cite{Cottam1986,Fleury1968}. In fact all these experiments performed in a huge variety of compounds demonstrated that the energy of the two-magnon mode is determined by the exchange energy and by a minor correction due to magnon-magnon interactions. Since the phase $\varphi_\mathbf{k}\approx\varphi\equiv \delta \omega_\mathrm{R}\tau_\mathrm{pulse}$ and the magnon frequency $\omega_\mathbf{k}$ show small dispersion, the contributions from all the modes to Eq. (\ref{eq_correlations}) are almost coherent. Summing the amplitudes we obtain the following expression for the longitudinal correlations

\begin{align}\label{eq_correlations_approximate}
\nonumber &\sum_{j,\boldsymbol{\delta}}\langle \hat{S}_j^z \hat{S}_{j+\boldsymbol{\delta}}^z \rangle=-\frac{1}{2}z_NNS^2+ z_NS \Delta L(0) +\\
&+z_NS4\pi\left(\frac{2\Delta ka}{3\pi}\right)^3\langle\hat{K}^z_\mathrm{{k_R}}\rangle e^{-t/\tau_\mathrm{d}}\sin(2\Omega t-\varphi),
\end{align}

\noindent and for the time-dependent longitudinal N\'eel component

\begin{eqnarray}\label{eq_L_z_approximate}
\delta L(t)= &\delta\omega_\mathrm{R}&\tau_\mathrm{pulse}4\pi\left(\frac{2\Delta ka}{3\pi}\right)^3\langle\hat{K}^z_\mathrm{{k_R}}\rangle e^{-t/\tau_\mathrm{d}}\cdot\\ \nonumber
&\cdot&\sin(2\Omega t-\delta \omega_\mathrm{R}\tau_\mathrm{pulse}).
\end{eqnarray}

\noindent where $\langle\hat{K}^z_\mathrm{{k_R}}\rangle = 1/2 + n_{\mathrm{k_R}}$, meaning that the expectation value of the K-operator is evaluated in the vicinity of the $R$-point. Here we introduce the decoherence time $\tau_\mathrm{d}(\Delta k)$ as a phenomenological parameter. \footnote{Note that this term has to be interpreted as an ensemble decoherence and not like a dephasing of the individual magnon modes involved in the 2M-process.~\cite{Bossini2016} The parameter $\tau_\mathrm{d}(\Delta k)$ represents the damping of the oscillations observable in a pump-probe experiment.~\cite{Bossini2016} An accurate calculation of $\tau_\mathrm{d}(\Delta k)$ goes beyond the harmonic approximation and thus the scope of our paper.}

Within the realm of the harmonic approximation, it holds that $n\ll1$. Moreover, since $\hbar\omega_R>T_N$, the thermal excitation of the femto-nanomagnons is very small in any temperature regime investigated. Hence we can conclude that the main contribution to $\langle\hat{K}^z_\mathrm{{k_R}}\rangle$ stems from the population of magnons due to quantum fluctuations, which are already present even in the ground state. It is important to underline that {the} experimental approach {introduced below} does not directly probe the ground state fluctuations themselves. In fact, the latter have a random phase-relation, forbidding the realization of a macroscopic coherent ensemble response which is, on the other hand, triggered by the photo-excitaton. In the all-optical experiments presented here and in the literature,~\cite{Bossini2016,Bossini:2016jmbaca,Bossini:2017bo} only the macroscopic ensemble dynamics with a uniquely defined phase can be detected.
	
\section{Methods and Materials}
\label{section:M&M}

\subsection{Magneto-optical pump-probe set-up}

For our experiments we used a regeneratively amplified mode-locked Ti:sapphire system delivering 100 fs pulses with central photon energy equal to 1.55 eV. The average power is 4 W and the repetition rate is 2 kHz. A 500 mW fraction of the laser output is used to drive two non-collinear Optical Parametric Amplifiers (NOPAs) operating in two different spectral ranges.~\cite{Brida2010} Both NOPAs are pumped by the second harmonic of the laser (i.e. 3.1 eV) and seeded by the white-light continuum produced by focusing the 1.55 eV beam into a sapphire plate. The amplified pulse from the first NOPA, which initiates the dynamics (pump), has a spectrum spanning the 2.45-1.75 eV range and is compressed to nearly transform-limited duration (i.e. 8 fs) by a pair of custom-made chirped mirrors. The amplified pulse, generated by the second NOPA (probe), covers the frequency range between 1.5 eV and 1.18 eV and is compressed to nearly  transform-limited duration (i.e. 13 fs) by a couple of fused silica prisms. 
The temporal resolution of the setup has been characterized by the cross-correlation frequency-resolved optical gating (XFROG) technique and was below 20 fs.\cite{DalConte2015a} The pump and probe beams were focused on the sample by a spherical mirror down to approximately 100 $\mu$m and 70 $\mu$m spot sizes, respectively. It is important to note that while the pump beam impinged on the sample surface at normal incidence (see Fig. \ref{fig:CrystalStructure}), the probe beam propagated at an angle ($< 10 ^{\circ}$) with respect to the pump \textcolor{black}{and with electric field close to the in-plane axes of the samples}. Consequently, in the case of K$_2$NiF$_4$ (see Fig. \ref{fig:CrystalStructure}(b)) a non-vanishing component of the probe beam propagates at an angle with the $c$-axis. The measurements on KNiF$_3$ were performed at a minimum temperature of 77 K in a liquid nitrogen cryostat. The high temporal resolution is preserved by using a very thin (200 $\mu$m) fused silica window as optical access to the liquid-nitrogen-cooled-cryostat. The experiments on K$_2$NiF$_4$ required liquid helium cooling, given the lower N\'{e}el temperature (T$_N$= 96 K versus T$_N$= 246 K in KNiF$_3$). The optical windows of the liquid-helium-cooled cryostat were 1 mm-thick sapphire plates. We pre-compressed the laser pulses by changing the optical path through the fused silica prisms, in order to preserve the superior time-resolution of our apparatus. The temperature of the samples was monitored in both cases by a thermocouple placed on the sample holder.

	\begin{figure}[h!]
	\center
	\includegraphics[]{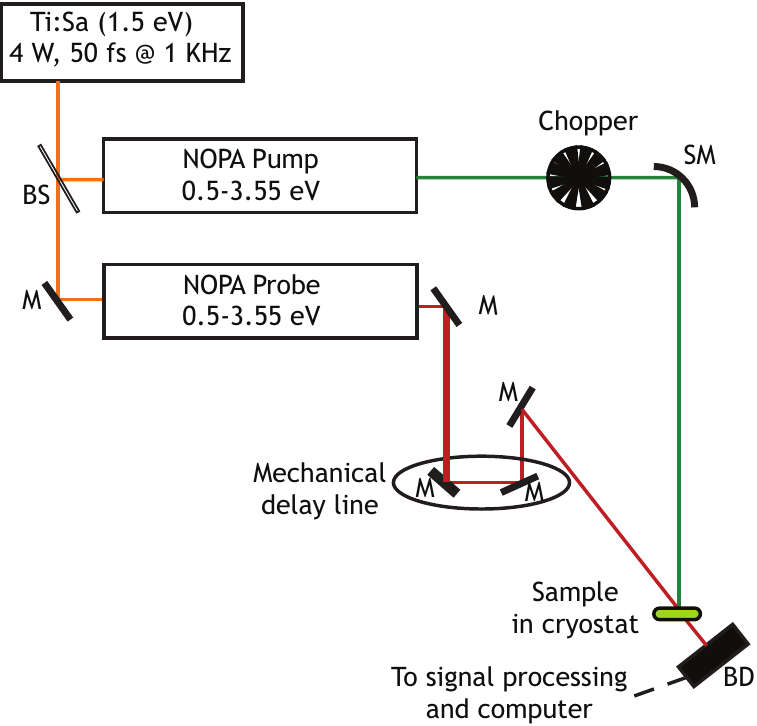}
	\caption{\footnotesize{Schematic representation of our set-up. The main components are shown: mirrors (M), spherical mirrors (SM), beam splitters (BS) non-collinear Optical Parametric Amplifiers (NOPA) and the balanced detector (BD). The polarization of the beams was linear and oriented along directions parallel to the crystal axes by means of half-waveplates and polarizers (not-shown).}}
	\label{fig:set-up}
	\end{figure}

After interaction with the sample the linearly polarized probe beam was focused sent to a balanced detection setup to measure the polarization rotation. Note that in Ref.~\cite{Bossini2014}, the detection was based on the measurements of the ellipticity.~\cite{Smolenski1976} To achieve it, an additional quarter-wave plate had to be placed between the sample and the detector. In the present experiments the quarter-wave plate was removed from the scheme and the rotation of the polarization was detected. The linearly polarized transmitted probe is split by a Wollaston prism into two orthogonal linearly polarized beams and focused on a couple of balanced photodiodes. The Wollaston prism is rotated in order to equalize the probe intensities on the two photodiodes. The pump-induced imbalance of the signal registered by the two photodiodes is measured by a lock-in amplifier which is locked to the modulation frequency of the pump beam (i.e. 1 kHz). \textcolor{black}{A schematic representation of the experimental set-up is reported in Fig. \ref{fig:set-up}. Our apparatus was able to detect rotations of the polarization down to 80 $\mu$deg. We did not employ a probe linearly polarized at 45$^{\circ}$ with respect to the pump because the sample birefringence perturbs our sensitive polarization rotation scheme.} \textcolor{black}{Although the first detection of the dynamics of two-magnon mode in antiferromagnets MnFe$_2$ and FeF$_2$ was based on time-resolved measurements of differential transmissivity, a more recent work employed the method of time-resolved polarization rotation measurements~\cite{Bossini2016}. Here we briefly review the physics of the probing mechanisms of the spins dynamics in antiferromagnets. In a linear light-matter interaction regime, the response of the media to the illumination is described in terms of dielectric permittivity tensor $\epsilon^{ij}$.  If in an otherwise isotropic medium (i.e. $\epsilon^{xx} = \epsilon^{yy} = \epsilon^{zz}$) the spin correlation function $\langle \hat{S}_i^{\Uparrow} \hat{S}_j^{\Downarrow} \rangle$ experiences a modification, such variation can be detected by optical methods due to a contribution to the symmetric part of the dielectric permittivity~\cite{Ferre1984} $\epsilon^{\lambda \nu}_s = \epsilon^{\nu \lambda }_s$}

	\begin{align}
	%\nonumber
	\delta \epsilon^{\lambda \nu}_{s} = \sum_{ij} \sum_{\gamma \delta} \rho^{\lambda \nu \gamma \delta} \langle \hat{S}_{i}^{\gamma \Uparrow} \hat{S}_{j}^{\delta \Downarrow} \rangle,
	\label{eq_ALD}
	\end{align}

\noindent where $\rho^{\lambda \nu \gamma \delta}$ is a phenomenological polar forth-rank tensor, ${ij}$ describe lattice sites and ${\lambda \nu \gamma \delta}$ are spatial coordinate indices. This contribution affects the absorption and refraction coefficients of the material due to isotropic contribution to the dielectric permittivity (i.e. $\delta \epsilon^{xx} = \delta \epsilon^{yy} = \delta \epsilon^{zz}$), thus modifying the intensity, reflected, absorbed and transmitted light beams, as reported~\cite{Zhao2006,Zhao2004} . An emergence of anisotropic contributions ($\delta \epsilon^{xx} \neq \delta \epsilon^{yy} \neq \delta \epsilon^{zz}$) would result in different absorption and refraction of a light beam linearly polarized along the $x$, $y$ and $z$-axis, respectively. Let us consider the propagation of a light beam along the $z$-axis. If $\Re{\epsilon^{xx}} \neq \Re{\epsilon^{yy}}$, the intensities of reflected beams polarized along the $x$- and $y$ axes respectively are different. On the other hand, in case $\Im{\epsilon^{xx}} \neq \Im{\epsilon^{yy}}$, the absorption experienced by beams polarized along the $x$- and $y-$axis respectively, and consequently the transmitted intensities, differ. As reported in the literature~\cite{Saidl2017}, this inequality results in a polarization rotation of the probe beam which is proportional to the modification of order parameter or, as in our case, of the spin correlation function $\langle \hat{S}_i^{\Uparrow} \hat{S}_j^{\Downarrow} \rangle$.

In optically anisotropic media $\epsilon^{xx} \neq \epsilon^{yy} \neq \epsilon^{zz}$ the optical detection of the dynamics of the spin correlation function $\langle \hat{S}_i^{\Uparrow} \hat{S}_j^{\Downarrow} \rangle$ is more complex. In fact, even if the spin correlation function contributed isotropically to the dielectric permittivity $\delta \epsilon^{xx} = \delta \epsilon^{yy} = \delta \epsilon^{zz}$, it would follow that $\epsilon^{xx} / \epsilon^{yy} \neq (\epsilon^{xx} + \delta \epsilon^{xx})/(\epsilon^{yy} + \delta \epsilon^{yy})$ and the dynamics of $\langle \hat{S}_i^{\Uparrow} \hat{S}_j^{\Downarrow} \rangle$ could still generate dynamics of the polarization rotation of the probe beam. Using the method of balanced detection, the intensity noise of laser sources can be greatly compensated and measurements of polarization rotation can be performed with an extreme sensitivity limited by the level of shot-noise. In order to achieve the highest possible sensitivity, we employed the technique of polarization rotation~\cite{Bossini2016}. We would like also to observe that since the response originates from the unbalance of $\epsilon^{xx}$ and $\epsilon^{yy}$ a probe beam polarized 45 degrees away from the crystal axes would maximize the signal. \textcolor{black}{However, this configuration did not provide best the signal-to-noise ratio, because of a strong increase of the background noise, whose origin was not investigated in details. Therefore we empirically selected the probe polarization resulting in the best signal-to-noise ratio. The best direction was found to be approximately parallel to one of the crystallographic axes. More precisely, in the case of KNiF$_3$ the polarization of the probe was approximately parallel to the $y(z)$-axis and in the case of K$_2$NiF$_4$ to the $x(y)$-axis. The degree of approximation is estimated to be of the order of 10$^{\circ}$}.
\textcolor{black}{
A drawback of our approach concerns the interpretation of the experimental results. In time-resolved studies of the dynamics of the $\langle \hat{S}_i^{\Uparrow} \hat{S}_j^{\Downarrow} \rangle$ four different scenarios can originate an anisotropy leading to the polarization rotation:}

\begin{enumerate}[(a)]
\item the dynamics of the spin correlation induces an isotropic contribution to the dielectric permittivity ($\delta \epsilon^{xx} = \delta \epsilon^{yy} = \delta \epsilon^{zz}$), but the medium is anisotropic in the unperturbed state ($\epsilon^{xx} \neq \epsilon^{yy} \neq \epsilon^{zz}$).
\item The dynamics of the spin correlation induces an anisotropic contribution to the dielectric permittivity ($\delta \epsilon^{xx} \neq \delta \epsilon^{yy} \neq \delta \epsilon^{zz}$), but  the medium is isotropic in the unperturbed state ($\epsilon^{xx} = \epsilon^{yy} = \epsilon^{zz}$).
\item The dynamics of the spin correlation induces an anisotropic contribution to the dielectric permittivity ($\delta \epsilon^{xx} \neq \delta \epsilon^{yy} \neq \delta \epsilon^{zz}$), and  the medium is anisotropic in the unperturbed state ($\epsilon^{xx} \neq \epsilon^{yy} \neq \epsilon^{zz}$).
\item Although the medium is isotropic in the unperturbed state ($\epsilon^{xx} = \epsilon^{yy} = \epsilon^{zz}$) and the spin correlation induced an isotropic contribution as well ($\delta \epsilon^{xx} = \delta \epsilon^{yy} = \delta \epsilon^{zz}$), the intense linearly polarized pump beam can induce an anisotropic transient linear birefringence of non-magnetic origin.
\end{enumerate}

\textcolor{black}{
In our experiment the dynamics of the spin correlation is induced by the intense pump pulse: as reported in our earlier work~\cite{Bossini2016}, the photo-induced dynamics of $\langle \hat{S}_i^{\Uparrow} \hat{S}_j^{\Downarrow} \rangle$ is a linear function of the pump intensity and provides an anisotropic contribution to the dielectric permittivity ($\delta \epsilon^{xx} \neq \delta \epsilon^{yy} \neq \delta \epsilon^{zz}$). Therefore the measured signal in the cases scenario (a) and (b) and (c) is expected to be linear with respect to the pump intensity, while only (b) and (c) are relevant to our experiment. More specifically, (b) can be ruled out since both materials here investigated are anisotropic before the photo-excitation. The main difference between the two compounds concerns the origin of the anisotropy: it arises from the crystal structure in K$_2$NiF$_4$ (being thus insensitive to the N\'eel temperature), while it has magnetic origin in KNiF$_3$ (sensitive to the N\'eel temperature). The last term, (d) can be neglected since it is expected to depend quadratically (or even with a higher degree of nonlinearity) on the intensity of the pump beam. This statement is motivated by the fact that photo-induced modification of the birefringence depends at the leading order linearly on the pump fluence, as the dynamics of the spin correlation function. Hence, the combined effect should display a non-linear dependence on the excitation fluence, in contrast with the observation reported in Section \ref{section:PumpFluenceDependence}.}

The SR spectrum of K$_2$NiF$_4$ was measured  in the backscattering geometry. The sample was excited by two different CW lasers, a diode with central photon energy of approximately 2.3 eV and a He-Ne source ($\approx$ 1.9 eV). The power of the incident radiation on the sample was 220 $\mu$W in the former case and 110 $\mu$W in the latter. The backscattered light was collected by a 10x objective (numerical aperture $\approx$ 0.25) and dispersed by a Horiba LabRam HR800 spectrometer. The detector was a cooled CCD camera, able to scan the Raman shift in the range from 200 to 700 cm$^{-1}$.  The sample was mounted on the cold finger of a liquid-nitrogen-cooled flow cryostat, held at a constant temperature of 70 K. The Raman shift was calibrated and the intensities were normalized by employing the 520 cm$^{-1}$ Si phonon peak measured under the same conditions.

\subsection{Materials}

	\begin{figure}[h!]
	\center
	\includegraphics{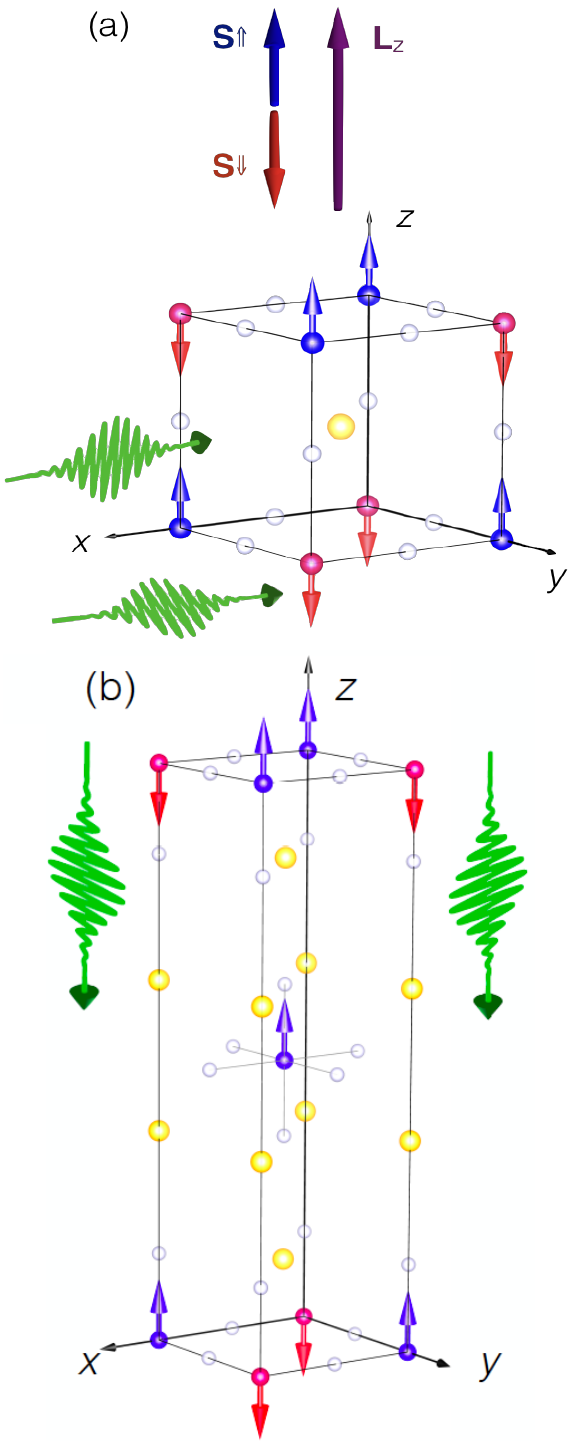}
	\caption{\footnotesize{Crystallographic and magnetic structures of KNiF$_3$ (a) and K$_2$NiF$_4$ (b). Note that the isolated spin between the NiF$_2$ layers in K$_2$NiF$_4$ is not relevant for the magnetic structure of this compound (see main text). We can thus consider these two samples as a 3D and a 2D Heisenberg antiferromagnet, respectively. The only exchange interaction relevant to the present study is between nearest-neighbor spins, which are located on two different ionic sites  and belong to oppositely oriented magnetic sublattices. The total spins of each sublattice, $\mathbf{S}^{\Uparrow}$ and $\mathbf{S}^{\Downarrow}$, are represented: they are obtained by summing all the magnetic moments belonging to the $\Uparrow$ and $\Downarrow$ sublattices, respectively. The magnetization ($\mathbf{M}$) and the antiferromagnetic vector ($\mathbf{L}$) are defined: while the former vanishes, the latter is the order parameter of an antiferromagnet. The propagation directions and the polarizations of the pump pulses employed during the experiments are represented by the green pulses in the figure.}}
	\label{fig:CrystalStructure}
	\end{figure}

We investigated two dielectric collinear antiferromagnets: the cubic KNiF$_3$ and the layer-structured (i.e. 2D) K$_2$NiF$_4$. Our KNiF$_3$ sample was a 340 $\mu$m thick (100) single crystal, which has a perovskite crystal structure. Two equivalent Ni$^{2+}$ sublattices are antiferromagnetically coupled below the N\'{e}el temperature $T_\mathrm{N} = 246$ K.\cite{Bossini2014} In the paramagnetic phase KNiF$_3$ is described by the $m3m$ point group, while in the ordered phase it belongs to the $4/mmm$ group. The ultrafast dynamics of the short-wavelength magnons in this system has already been reported.~\cite{Bossini2016} Here we discuss the dependence of the signal on the temperature and on the polarization of the pump beam, in comparison with the results obtained for the uniaxial antiferromagnet.

The structure of K$_2$NiF$_4$ consists of antiferromagnetic planes of NiF$_2$ separated by KF planes, which is similar to the atomic arrangement of superconducting cuprates of La$_2$CuO$_4$ type (see Fig. \ref{fig:CrystalStructure}(b)). Our specimen is a  800 $\mu$m thick single crystal, cut perpendicular to the $c$-axis. This material orders at T$_N \approx 96$ K. Also K$_2$NiF$_4$ belongs to the $4/mmm$ group in the antiferromagnetic phase, where the orientation of the 4-fold axis is given by the orientation of antiferromagnetic vector. The dominant exchange interaction determines an antiparallel alignment of the Ni$^{2+}$ spins in the NiF$_2$ planes, via 180$^{\circ}$ Ni-F-Ni bonds.~\cite{Lines1967,Abdalian1988} Even neglecting the interplane exchange interaction between the Ni$^{2+}$ ions in the ordered planes and the isolated Ni$^{2+}$ ion between the planes (which is at least one order of magnitude weaker than the in-plane exchange coupling~\cite{Lines1967}), the bulk properties of this compound are properly described.

While for both these compounds the exchange interaction is taken into account by means of the nearest-neighbours Heisenberg interaction, the magnetocrystalline anisotropies strongly differ. In the case of KNiF$_3$ a very weak cubic magnetic anisotropy with positive sign of the anisotropy constant determines the alignment of spins along the [001], [010], and [100] axes.~\cite{Landau1984} \textcolor{black}{The size of the domains was reported to be on the mm-scale, so that the spot size of our focused laser beams (70-100 $\mu$m) allows to interrogate a single domain in this material~\cite{Safa1978}.}
On the other hand the sublattice magnetization in K$_2$NiF$_4$ is parallel to the $c$-axis, due a single-axis anisotropy.~\cite{Lines1967}

	\begin{figure}
	\center
	\includegraphics{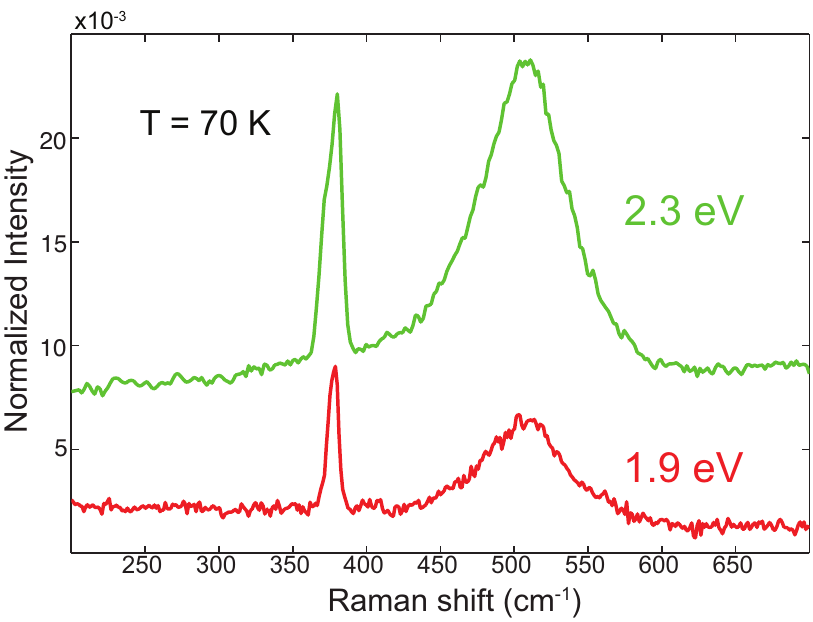}
	\caption{\footnotesize{The spontaneous Raman spectrum of K$_2$NiF$_4$, measured with 2.3 eV and 1.9 eV CW lasers. These photon energies are comparable with the ones employed in the pump-probe experiments. The power of the incident radiation on the sample was 220 $\mu$W for the 2.3 eV excitation, while it amounted to 110 $\mu$W when the photon energy was 1.9 eV. The temperature was 70 K.}}
	\label{fig:Raman}
	\end{figure}

Raman spectroscopy investigations revealed the features of the 2M-mode in K$_2$NiF$_4$.~\cite{Fleury1970,Chinn1971,Toms1974} At low temperature (10 K) the Raman shift is approximately 520 cm$^{-1}$, corresponding to $\nu \approx 15.6$ THz (period $\approx 65$ fs), while the linewidth (FWHM) is 100 cm$^{-1}$, from which a lifetime on the order of 330 fs is expected. The spectrum of this material shows also several Raman-active phonon modes~\cite{Toms1974}, in particular a collective vibration with frequency in the $11$ THz range ($\approx 380$ cm$^{-1}$). These observations have been confirmed by the measurement of the SR spectrum on our specimen of K$_2$NiF$_4$, reported in Fig. \ref{fig:Raman}. Although the long-range magnetic properties are dramatically different for these compounds,~\cite{Mermin1966,Lines1967} the experimental evidence concerning the magnons near the edges of the Brillouin zone in K$_2$NiF$_4$ are comparable to the case of KNiF$_3$, as discussed in section \ref{section:Tdep}. 

\section{Temperature Dependence of the femto-nanomagnons}
\label{section:Tdep}

\textcolor{black}{Differently from low-energy collective spin excitations with wavevector at the center of the Brillouin zone, the frequency of the 2M-mode does not soften upon approaching the N\'eel point. Moreover, spontaneous Raman experiments have detected a peak at the characteristic frequency of the 2M excitation even when the temperature was higher than T$_{\mbox{\footnotesize{N}}}$. This is common to basically all the antiferromagnets investigated~\cite{Cottam1986}. It was conventionally accepted that, the Raman signal is detected above the N\'eel temperature because short-range spin correlations} %with characteristic size larger than the wavelength of the paired magnons (..nm) can be present in antiferromagnets even above the Neel point.  
%contribute to the 2M-mode. 

In contrast with SR experiments, the time-domain observations of the spin dynamics induced by optically generating the 2M-mode have failed to reproduce this experimental trend in several materials.~\cite{Bossini2016,Zhao2006} To be more precise, although the temperature dependence of the frequency of the 2M-mode did not reveal any noticeable softening,~\cite{Bossini2016} the amplitude of the femto-nanomagnonic oscillations decreased upon a temperature increase and no signal has ever been observed above the N\'eel temperature.~\cite{Bossini2016,Zhao2006} Therefore it has been suggested that long-range spin correlation play "an important, if not essential role" for the 2M process,~\cite{Zhao2006} \textcolor{black}{pointing towards the possibility of a discrepancy in the results obtained employing the two different experimental approaches.} 

We study this open problem by measuring and comparing the temperature dependence of the spin dynamics observed in KNiF$_3$ and K$_2$NiF$_4$. \textcolor{black}{These materials have different magnetic anisotropies, therefore long-range spin-correlations, which presumably affect the 2M-process in these compounds, are also expected to be different.} Note that the optical spectrum of the two compounds is almost identical, except for the fine-structure splitting of some \textit{d-d} transitions, which is however too tiny to be resolved by broadband femtosecond laser pulses.~\cite{Abdalian1988} Consequently, a straightforward comparison of the results obtained with these two samples can be carried out only based on their magnetic properties.

	\begin{figure}
	\center
	\includegraphics{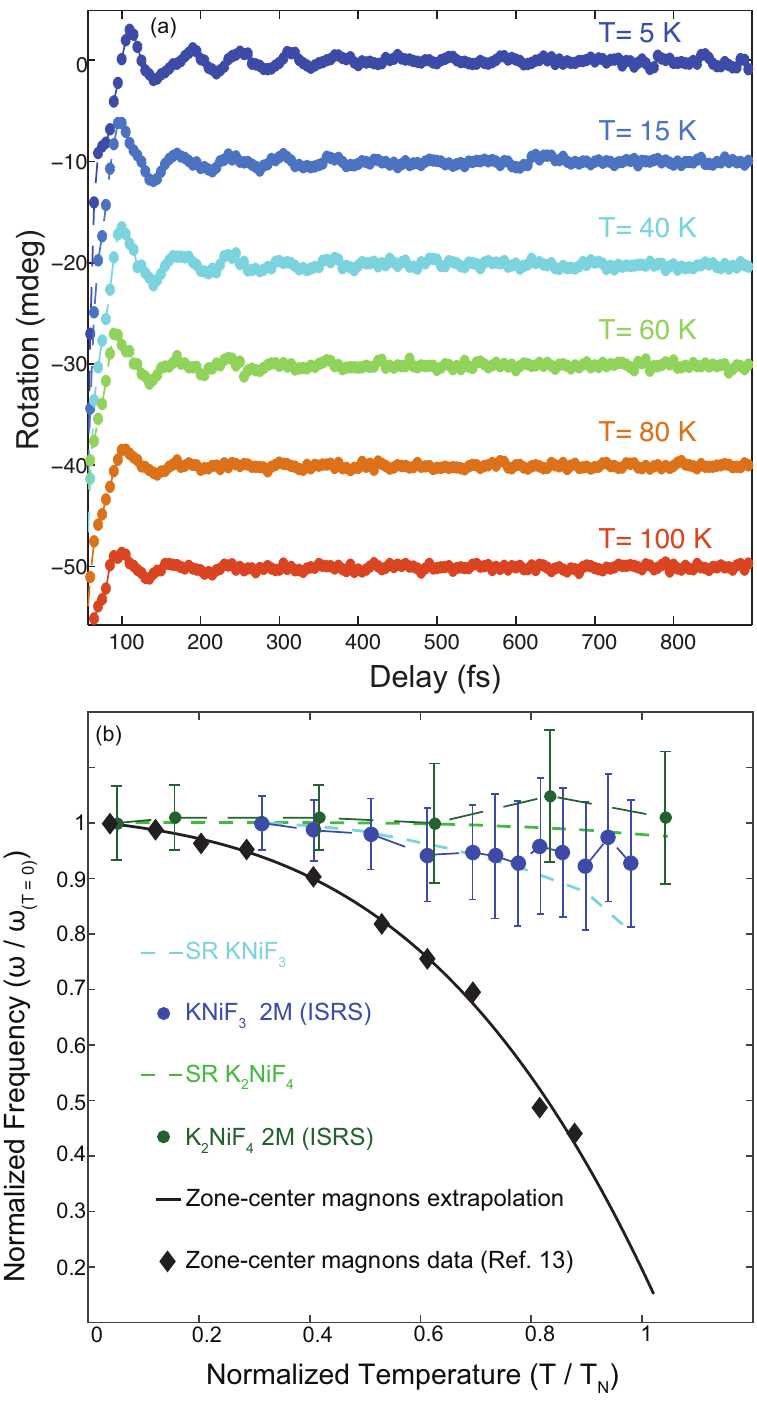}
	\caption{\footnotesize{Temperature dependence of normalized frequency of the two-magnon mode. (a) Spin dynamics as a function of the temperature measured in K$_2$NiF$_4$. The pump and probe photon energies were 1.9 eV and 1.3 eV respectively. The fluence was set to $\approx$ 4.5 mJ/cm$^2$ and the pump beam was circularly polarized. (b) The frequency of the oscillations is estimated from the Fourier transform of the data. Following a procedure typically employed in the Raman literature,~\cite{Toms1974} the frequency of the oscillations is normalized to the value of the frequency detected at the lowest temperature. Note that the minimum temperature achieved for the two samples is different: 80 K for KNiF$_3$ and 5 K in K$_2$NiF$_4$. The errorbars are defined as half-width-half maximum of the two-magnon spectrum. The dashed lines represent the result of SR scattering experiments on KNiF$_3$~\cite{Bossini2016} and K$_2$NiF$_4$.~\cite{Toms1974} The black diamonds represent experimental results of the frequency of the low-energy magnons KNiF$_3$ from reference.~\cite{Bossini2014} The black continuous line was calculated using these results, to demonstrate the characteristic softening of the long-wavelength spin waves near the N\'eel point.}}
	\label{fig:TdepFreq}
	\end{figure}

Figure \ref{fig:TdepFreq}(a) reports the temperature dependence of the spin dynamics optically excited and detected in the time-domain in K$_2$NiF$_4$. The corresponding measurements for KNiF$_3$ are reported in Reference~\cite{Bossini2016}. Figure \ref{fig:TdepFreq}(b) reports the temperature-dependence of the frequency of the 2M-mode observed in the time-domain in KNiF$_3$ (\textcolor{black}{blue} dots) and K$_2$NiF$_4$ (green dots). In the latter case, it was possible to observe oscillations also when the temperature was set slightly above the N\'eel point. Note that our results are consistent with experimental investigations of the temperature dependence of the 2M-mode by means of SR spectroscopy (see dashed lines in Fig. \ref{fig:TdepFreq}(b)). On the same panel we plot the typical temperature dependence of the frequency of $\mathbf{k} \approx 0$ magnons in antiferromagnets,~\cite{Bossini2014} exhibiting the characteristic softening as the temperature of the sample approaches the N\'eel point. It is important to observe that also the long-wavelength magnons in K$_2$NiF$_4$ soften at T$_\mathrm{N}$, as demonstrated experimentally.~\cite{Birgeneau1970} It is evident that the temperature dependence of the 2M-frequency of both samples does not display any softening. 

According to the well-established spin wave theory, the frequency of magnons near the centre of the Brillouin zone is defined by both the effective magneto-crystalline anisotropy field and the exchange interaction.~\cite{Cottam1986} On the other hand, the exchange energy only is relevant for the frequency of magnons near the edges of the Brillouin zone, which are the spin excitations involved in the 2M-mode. The absence of any softening in the data-set shown in Fig. \ref{fig:TdepFreq}(b) implies that the 2M-excitation itself is not affected by the long-range order. 

Thus the question naturally arises: why is the signal not observed well-above the N\'eel point in the time-domain as it occurs in the frequency-domain Raman spectroscopy? In our opinion, the explanation does not lie in the difference between generation of magnons via stimulated and spontaneous Raman scattering, but in the different detection of the 2M-mode in frequency-resolved and time-resolved Raman experiments. In particular, even above the N\'eel point, where the net magnetic order parameter is vanishing, short-range spin-spin correlations are still present and a light beam can still scatter on 2M-excitations. Relying on a spectral analysis of the scattered light beam, experimental signatures of the 2M-excitations in the paramagnetic phase can be detected.~\cite{Cottam1986} On the other hand, in the case of the time-resolved approach the situation is different:  \textcolor{black}{the detection of} the time-dependence of the spin correlation \textcolor{black}{is phase-sensitive. Once the long-range magnetic order is lost above the N\'eel temperature, the oscillations of the magnetic ensemble do not have the same phase, and thus average out.}
As a result, a time-resolved detection of coherent magnons at the edge of the Brillouin zone is not possible. We observe that even if a single pump-probe trace revealing the 2M-mode was observed when the temperature was set slightly above the N\'eel point in K$_2$NiF$_4$ (see Fig. \ref{fig:TdepFreq}(a)), it does not mean that the whole sample volume probed in the measurement was in the paramagnetic phase. In fact, in the case of wide band-gap dielectric materials with poor thermal conductivity a mismatch between the temperature set value and the material temperature on the order of several Kelvin typically occurs.
%is equivalent to the dynamics of the macrospin N\'eel vector~\cite{Bossini2016}. Above the N\'eel temperature the $\mathbf{L}$ vector is zero by definition and hence it cannot oscillate. 

Having established that only the short-range spin-spin correlations and zone-edges magnons are relevant to the femto-nanomagnonics, we can also conclude that, in this particular dynamical regime, the magnetocrystalline anisotropy does not play an important role, since the properties of zone-edge magnons are dominated by the exchange interaction. However, the long-range magnetic properties are important for the magneto-optical detection of the zone-edge magnons in time-resolved experiment.

\section{Pump Polarization dependence}
\label{section:PumpPolDependence}

\textcolor{black}{The effect of the pump-beam polarization on the amplitude and the phase of the oscillations of the antiferromagnetic vector $\mathbf{L}$ was reported for KNiF$_3$,~\cite{Bossini2016} but the origin has not been discussed in the literature yet. Here we explore the pump-polarization dependence of the femto-nanomagnonic signal in materials with very similar optical properties, but substantially different spin structures (i.e. KNiF$_3$ and K$_2$NiF$_4$).} 
	
\begin{figure}
	\center
	\includegraphics{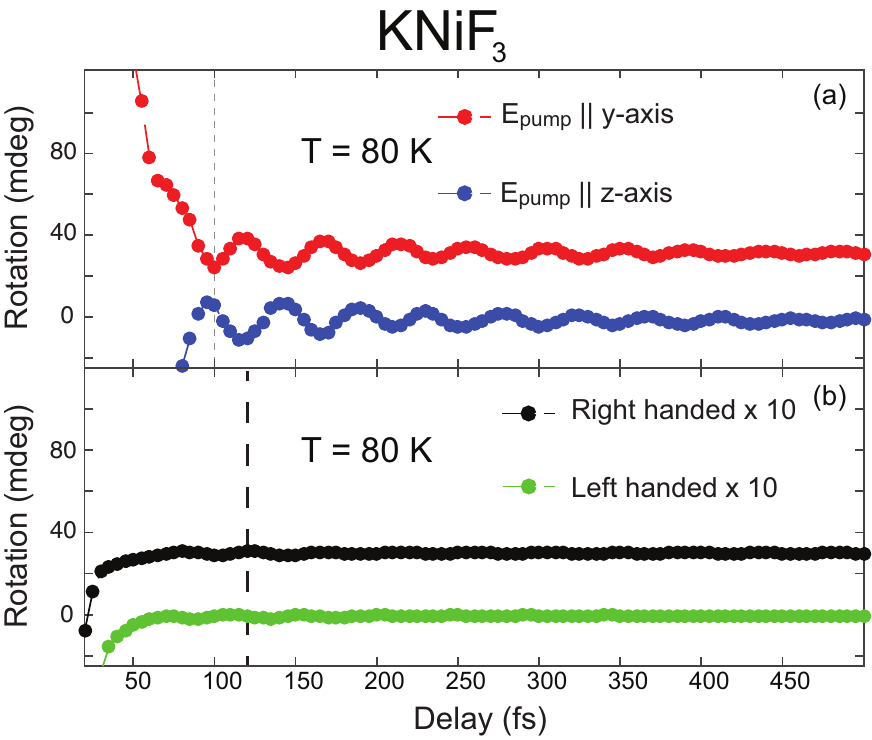}
	\caption{\footnotesize{ISRS measurements in KNiF$_3$ using pump pulses with different linear (a) and circular (b) polarization. The measurements were performed at 80 K. The pump and probe photon energies were 2.2 eV and 1.3 eV respectively. The grey and black dashed lines highlight the phase relation among the different time traces in the case of linearly or circularly polarized pump beam, respectively. The experimental geometry is shown in Fig. \ref{fig:CrystalStructure}(a). (a) The excitation beam was linearly polarized and the pump fluence was on the order of 10 mJ/cm$^2$. (b) The data (multiplied by a factor 10) here shown were obtained by employing circularly polarized pump beams. The fluence was on the order of 20 mJ/cm$^2$.}}
	\label{fig:PumpPolDepKNiF3}
	\end{figure}

Let us first consider KNiF$_3$. The ISRS dynamics reported in Fig. \ref{fig:PumpPolDepKNiF3}(a) shows that rotating the electric field of the linearly polarized pump beam from one axis to the other results in a $\pi$-shift of the phase of the oscillations. 
%This result is in agreement with phenomenological considerations of the symmetry of the light-spin interaction term in a thermodynamical potential (see Eq.  (\ref{eq_potential_tetragonal_final}) and the subsequent discussion in section \ref{section:Theory}).
\textcolor{black}{Since we employed laser beams linearly polarized along both the $y$ and $z$ axes, the configuration shown in Fig. \ref{fig:CrystalStructure}(a) is a proper representation of the experiment: regardless of which domain is contributing to our signal ($y$ or $z$ domain), we explored both the conditions of electric field parallel and perpendicular to spins.} This is fully consistent with the well-known magnetic configuration of KNiF$_3$, which consists of spins aligned along directions parallel to the crystallographic axes.~\cite{Bossini2014}
	
We analyze now the excitation induced employing circularly polarized light, for which the $z$- and $y$-components of the electric field have a phase difference equal to $\pi/2$. Note that for this polarization state the following property holds: $E_yE_z^{\ast}=-E_zE_y^{\ast}$. Making use of this properties it is possible to show that the light-matter interaction vanishes for both helicities (following an approach reported in the literature~\cite{Bossini2016}) entailing that the excitation of the 2M-mode is forbidden for circularly polarized laser pulses.
This, in principle, is not surprising. In fact a circularly polarized optical beam accesses the antisymmetric components of the dielectric tensor, which are proportional to the odd powers in spin~\cite{Zvezdin1997,Landau1984}. The 2M scattering is a process quadratically dependent on the spin and it is therefore described by the symmetric components of the dielectric tensor.\cite{Cottam1986,Fleury1968} Therefore, a purely circularly polarized beam cannot generate coherent femto-nanomagnons. However, a tiny although detectable magnonic oscillations were observed by pumping KNiF$_3$ with a circularly polarized optical beam (see Fig. \ref{fig:PumpPolDepKNiF3}(b)). This apparent discrepancy with the symmetry analysis is due to an imperfect polarization of the pump beam. The superior time resolution of our experiment is obtained by using broadband laser pulses (FWHM $\approx$ 40 nm). The broadband quarter waveplate employed to generate the circular polarization state has a slightly different retardation for each spectral component of our pulses. Consequently, the polarization state of the beam after the waveplate is elliptical.
Since an elliptically polarized beam can be described as the sum of a circularly and a linearly polarized beam, we ascribe the oscillations observed in Fig. \ref{fig:PumpPolDepKNiF3}(b) to the residual linear component.  \textcolor{black}{The suggested phenomenological model in Section \ref{sub:PumpPolDepTheo} predicts the polarization dependence of the laser-induced two-magnon oscillations in KNiF$_3$; moreover we would also like to observe that, even in a fully isotropic medium, linearly polarized light $E_jE_j$ couples to specific components of spins $\langle S_jS_j\rangle$. A rotation of the polarization of the pump over 90 degrees would corresponds to the excitation of different spin components. This will result in a sign change of the polarization rotation induced by the ALD  (see Supplementary Materials of Ref. ~\cite{Saidl2017}). To fully shed light on the mechanism generating the polarization dependence, experiments with different orientations of the polarizations of the pump and the probe beams need to be performed.} This is a subject of future studies. However, independently of the outcome of the future suggested experiments,  this discussion seems to be irrelevant in K$_2$NiF$_4$.
	\begin{figure}
	\center
	\includegraphics{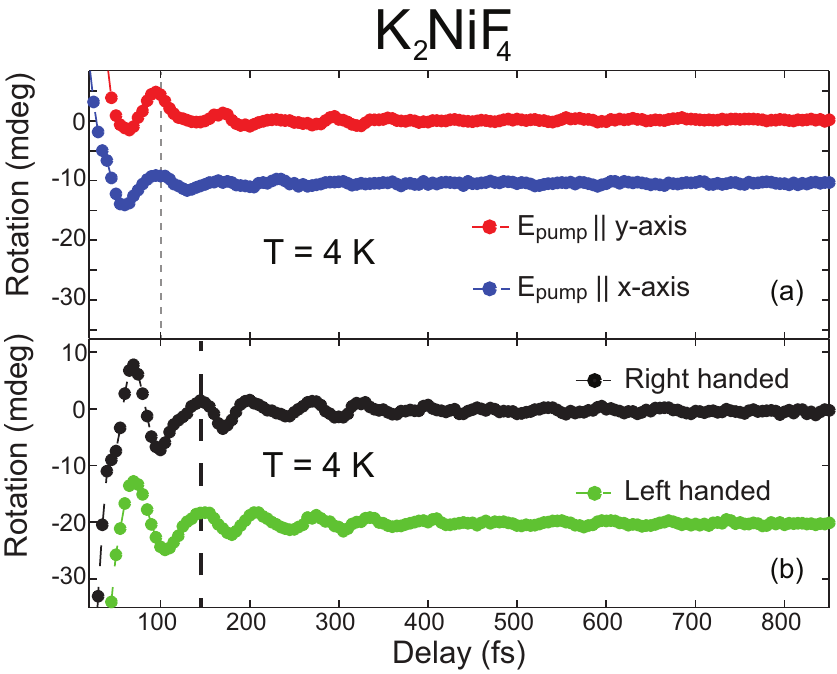}
	\caption{\footnotesize{The measurements of K$_2$NiF$_4$ were performed at 4 K. The pump and probe photon energies were 1.9 eV and 1.3 eV respectively. The grey and black dashed lines highlight the phase relation among the different time traces in the case of linearly or circularly polarized pump beam, respectively. The experimental geometry is shown in Fig. \ref{fig:CrystalStructure}(b). (a) The excitation beam was linearly polarized and the pump fluence was $\approx$ 3 mJ/cm$^2$. (b) The data here shown were obtained by employing circularly polarized pump beams. The fluence was $\approx$ 4 mJ/cm$^2$.}}
	\label{fig:PumpPolDepK2NiF4}
	\end{figure}
The ISRS results for K$_2$NiF$_4$ displayed in Fig. \ref{fig:PumpPolDepK2NiF4} are significantly different from the KNiF$_3$ case: the phase of the oscillations never changes regardless of the polarization of the pump beam. \textcolor{black}{We note that the measurements performed with circularly and linearly polarized light are two different experiments, since different waveplates are employed and thus the optical alignment differs as well. Consequently, experiments with linearly and circularly polarized light cannot be compared.}

\textcolor{black}{From Eq.~(\ref{eq_effective field}) it follows that the symmetry allows different values of the generalized force $H_{\mu}$ for the parallel ($e_\|=1$, $e_\perp=0$)  and perpendicular ($e_\|=0$, $e_\perp=1$) configurations. Considering Eq.~(\ref{eq_effective field}) we can then formulate the following predictions:
\begin{itemize}
	\item if light propagates along the N\'eel vector, as in the case of K$_2$NiF$_4$, $e_\|=0$, $\Delta L\propto H_\mu\propto \left(\chi_{13}-\chi_{11}\right)$ and does not depend upon the direction of light polarization. 
	\item If light propagates perpendicularly to the N\'eel vector, as in the case of KNiF$_3$, both $e_\|$ and $e_\perp$ components of the polarization vector could be nonzero. The difference between two orthogonal polarization states $\mathbf{E}\|(e_\|, e_\perp)$ and $\mathbf{E}\|(- e_\perp,e_\|)$ is $\propto (\chi_{33}-\chi_{31}-\chi_{13}+\chi_{11})(e^2_\|-e^2_\perp)$. It is maximal when light is polarized parallel/perpendicular to the N\'eel vector.
\end{itemize}
}	
In accordance with the symmetry analysis no polarization dependence was observed in KNi$_2$F$_4$ if the exciting laser pump beam propagates along the antiferromagnetic vector. It may be surprising that the oscillations are observed at all. According to our model the antiferromagnetic order is probed due to ALD, which in the case of K$_2$NiF$_4$ should be zero, if light propagates along the antiferromagnetic vector. The apparent contradiction is explained by the fact that during the experiment the probe and the pump beams were impinging on the sample at different angles. Therefore, if the pump was propagating nearly along the antiferromagnetic vector, the wavevector of the probe was at an angle to \textbf{L} and magnetic linear dichroism was allowed. To verify this hypothesis experiments with different orientations of the crystal are necessary. Such experiments are subject of future studies.

\textcolor{black}{Moreover, analogously to the case of KNiF$_3$, symmetry arguments determine that a purely circularly polarized beam cannot generate coherent femto-nanomagnons in K$_2$NiF$_4$ as well. However, in addition to the aforementioned observation concerning the purity of the polarization state of our laser beams, we would like to remind that static magneto-optical effects induce further distortions of the polarization. Considering the remarkable thicknesses of our samples (in particular  of K$_2$NiF$_4$), the magneto-optical effects, which are proportional to the propagation distance of light in a magnetic material, play a non-negligible role in modifying the polarization of the laser beams. Therefore a magnonic signal was detected also by illuminating the sample with circularly polarized laser pulses (Fig. \ref{fig:PumpPolDepK2NiF4}(b)).}

\section{Pump Fluence dependence}
\label{section:PumpFluenceDependence}
	\begin{figure}
	\center
	\includegraphics{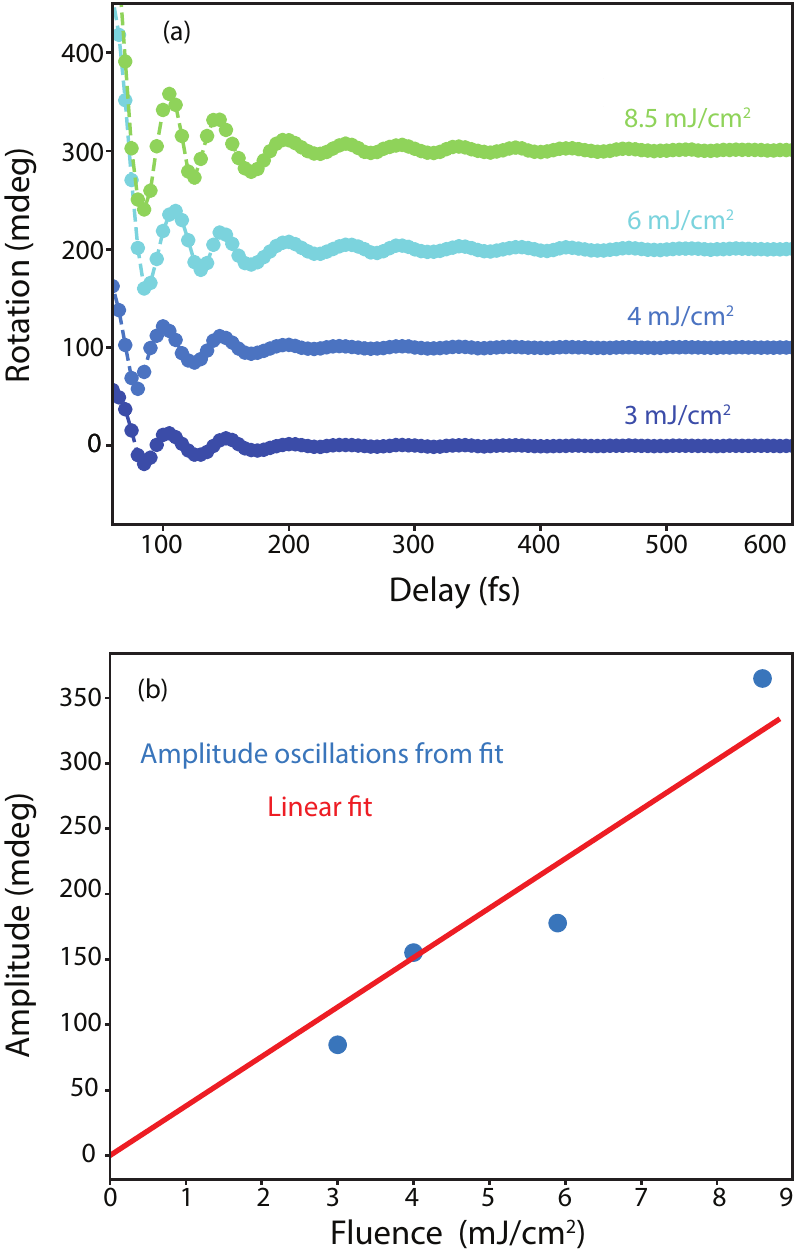}
	\caption{\footnotesize{The measurements were performed on KNiF$_3$ at 77 K. (a) Time-dependence of the photo-induced spin dynamics for several pump fluences. The time-traces are shifted vertically for presentation purposes. (b) \textcolor{black}{The time-traces were fitted with a damped harmonic function in order to retrieve the amplitude of the mode.} The dependence of the amplitude of the oscillations on the excitation fluence can be described by a linear function (obtained via fitting); most importantly no trace of a quadratic trend is visible}.}
	\label{fig:FlDep}
	\end{figure}

\textcolor{black}{The dependence of the spin dynamics on the excitation fluence was investigated. This experiment was performed on KNiF$_3$ at liquid nitrogen temperature. The pump and probe photon energies were 1.9 eV and 1.3 eV respectively; both beam were linearly polarized. The results, shown in Fig. \ref{fig:FlDep}, reveal that the amplitude of the magnonic signal is enhanced as the excitation intensity increases. Moreover, \textcolor{black}{we estimated the amplitude of the magnonic signal by fitting the time-traces with a damped harmonic function. The result of this procedure is shown in Fig. \ref{fig:FlDep}(b), in which the fluence dependence of the amplitude of the oscillations is shown. The obtained trend is consistent with a linear function. Importantly, no clear signatures of quadratic or even higher-order effects are visible. This provides evidence that a classical description of two-magnon dynamics is not adequate, since it predicts spin oscillations scaling quadratically with the pump intensity, as discussed in Section~\ref{section:Theory} and reported~\cite{Baierl2016}.} Consequently, even the magneto-optical detection has to rely on effects linearly proportional to the intensity of light, as reported in Eq.(\ref{eq_ALD}).}

\section{Conclusions and perspectives}
\label{section:conclusions}

The present experimental and theoretical results provide a comprehensive description of the spin dynamics triggered by the ISRS excitation of femto-nanomagnons. We have experimentally demonstrated that, despite the vanishing wavevector exchanged during the light-matter interaction, only spin-excitations near the edges of the Brillouin zone are actually involved. This conclusion is confirmed even further noticing that consistent results were obtained in antiferromagnets with different magneto-crystalline anisotropy (cubic-KNiF$_3$ and uniaxial K$_2$NiF$_4$). The data allowed also to derive a criterion for the experimental conditions required to achieve the phase control of the magnons. The femto-nanomagnonic regime here reported is a unique case of light-induced spin dynamics which cannot be interpreted in the realm of classical physics. The coherent longitudinal oscillations of the antiferromagnetic vector triggered by laser pulses demand a quantum mechanical approach. Formulating a microscopic description in terms of perturbations of exchange interactions and magnon pair operators, we employ coherent states for these operators to provide an effectively macroscopic equation of motion beyond thermodynamics for the coherent longitudinal dynamics of the antiferromagnetic vector. Our model predicts that the pairs of magnons involved in this process are entangled, which may be verified in a non-local experimental scheme. This fact, combined with the achievement of the conditions required to manipulate the phase of the magnonic oscillations, may pave the way to a concept of femtosecond manipulation of the entanglement in solid state compounds aimed at ultrafast quantum information technology. We want to stress that no classical counterpart of this peculiar spin dynamical regime exists.

Although our experiments have been limited to model systems, the approach can be extended to several classes of materials. For instance, pumping the 2M-mode in ferrimagnets and weak antiferromagnets would address the fascinating question of the role played by the magnetization in the highest-possible frequency spin dynamics. Time-resolved experiments in such materials could photo-induce and manipulate the ultimately fastest dynamics of the angular momentum. It is also worth to mention that femto-nanomagnons have been hitherto studied exclusively in collinear spin structures. Different systems, especially spin textures with topological objects like bubbles, skyrmions and domain walls with a size comparable with the wavelength of the femto-nanomagnons, could be dramatically affected by the photo-excitation of magnons near the edges of the Brillouin zone. Another enticing perspective consists in exciting resonantly the femto-nanomagnons. Although a direct ultrafast excitation of the 2M-mode is far from trivial and unexplored so far, a first demonstration of resonant pumping of magnons via a different mechanism has recently been reported to be able even to induce magnetic phase-transitions on the femtosecond timescale~\cite{Bossini2018}.

The research leading to this results was supported by the Japanese Society for Promotion of Science (JSPS) "Postdoctoral Fellowship for Overseas Researcher" No P16326, de Nederlandse Organisatie voor Wetenschappelijk Onderzoek (NWO), de Stichting voor Fundamenteel Onderzoek der Materie (FOM), ERC Advanced Grant 339813 (EXCHANGE), LASERLAB-EUROPE (grant agreements no. 284464 and no. 654148, EC's Seventh Framework Programme). R.V.P. acknowledges the support from the Russian Science Foundation, project no. 16-12-10456. J.H.M. acknowledges funding by the Nederlandse Organisatie voor Wetenschappelijk
Onderzoek (NWO) by a VENI grant, by the European Research Council
(ERC) Advanced Grant No. 338957 (FEMTO/NANO) and is part of the Shell-NWO/FOM-initiative "Computational sciences for energy research" of Shell and Chemical Sciences, Earth and Life Sciences, Physical Sciences, FOM and STW. H.G. and J.S. acknowledge support from Humboldt Foundation, EU ERC Advanced Grant No. 268066 and the Transregional Collaborative Research Center (SFB/TRR) 173 SPIN+X. G.C. acknowledges support by the European Union Horizon 2020 Programme under Grant Agreement No. 696656 Graphene Flagship. D. Bossini, S. Dal Conte, J. Mentink and H. Gomonay equally contributed to this work.

\appendix
\section{Appendix}
\label{a:magnonpair}

In this appendix we show how $H_0$ and $\delta H$ (Eqs.(\ref{H0}) and (\ref{dH}) of the main text) can be written in terms of magnon annihilation (creation) operators $\hat{\alpha}_\mathbf{k}\,(\hat{\alpha}^\dagger_\mathbf{k})$ and $\hat{\beta}_\mathbf{k}\,(\hat{\beta}^\dagger_\mathbf{k})$. Following the literature,~\cite{kittel1963,fazekas1999} our approach consists in computing weak deviations from the classical N\'eel state, which we express by introducing Holstein-Primakoff bosons for sublattice $A$ and $B$:

	\begin{align}
&S_{Ai}^{+}=\sqrt{2S}\left(1-\frac{a_i^\dagger a_i}{2S}\right)^{1/2}\!\!\!\!\!\!a_i, \qquad S_{Ai}^{-}=\sqrt{2S}a_i^\dagger\left(1-\frac{a_i^\dagger a_i}{2S}\right)^{1/2}\!\!\!\!\!\!,\nonumber \\
&S_{Ai}^z=S-a_i^\dagger a_i, \\
&S_{Bi}^{+}=\sqrt{2S}b_i^\dagger\left(1-\frac{b_i^\dagger b_i}{2S}\right)^{1/2}\!\!\!\!\!\!, \qquad S_{Bi}^{-}=\sqrt{2S}\left(1-\frac{b_i^\dagger b_i}{2S}\right)^{1/2}\!\!\!\!\!\!b_i,\nonumber \\
&S_{Bi}^z=-S+b_i^\dagger b_i. 
	\end{align}

\noindent Attempting to obtain magnon states we introduce the Fourier transforms:

	\begin{align}
a_\mathbf{k}^\dagger=\sqrt{\frac{2}{N}}\sum_i e^{-\text{i}\mathbf{k}\cdot\mathbf{R}_i}a^\dagger_i \qquad
a_i^\dagger=\sqrt{\frac{2}{N}}\sum_\mathbf{k}e^{\text{i}\mathbf{k}\cdot\mathbf{R}_i}a^\dagger_\mathbf{k},\\
b_\mathbf{k}^\dagger=\sqrt{\frac{2}{N}}\sum_i e^{-\text{i}\mathbf{k}\cdot\mathbf{R}_i}b^\dagger_i \qquad
b_i^\dagger=\sqrt{\frac{2}{N}}\sum_\mathbf{k}e^{\text{i}\mathbf{k}\cdot\mathbf{R}_i}b^\dagger_\mathbf{k}.
	\end{align}

\noindent Here the factor $\sqrt{2/N}$ appears since the magnetic Brillouin zone contains only $N/2$ $\mathbf{k}$-vectors, with $N$ the number of sites in the original lattice. After these transformations the harmonic part of $H_0$ becomes

	\begin{align}\label{H0harm}
H_0 \approx& -\frac{z_NJN}{2}S^2+ z_NJS\sum_\mathbf{k}\left[\gamma_\mathbf{k}(a_\mathbf{k} b_{-\mathbf{k}} + b_\mathbf{k}^\dagger a_{-\mathbf{k}}^\dagger) + (a_\mathbf{k}^\dagger a_\mathbf{k} + b_\mathbf{k}^\dagger b_\mathbf{k})\right]\nonumber, 
	\end{align}
	
\noindent where $\gamma_\mathbf{k}=\frac{1}{z_N}\sum_{\boldsymbol{\delta}}\exp{(i\mathbf{k}\cdot\boldsymbol{\delta})}$ depends on the geometry of the exchange bonds given the sum over nearest-neighbor bonds $\boldsymbol{\delta}$. Due to the nearest-neighbors coupling between the sublattices $A,B$, the operators $a_\mathbf{k},b_\mathbf{k}$ do not diagonalize $H_0$. The physical magnons therefore comprise superpositions of $a_\mathbf{k}$ and $b^\dagger_{-\mathbf{k}}$ as described by the Bogoliubov transformation:

	\begin{align}
a_\mathbf{k}=u_\mathbf{k}\alpha_\mathbf{k} + v_\mathbf{k}\beta_{-\mathbf{k}}^\dagger, \\
b_\mathbf{k}=u_\mathbf{k}\beta_\mathbf{k} + v_\mathbf{k}\alpha_{-\mathbf{k}}^\dagger. 
	\end{align}
	
\noindent Requiring the Bose commutation relations $[\alpha_{\mathbf{k}},\alpha^\dagger_{\mathbf{k}}]=1$, $[\beta_{\mathbf{k}},\beta^\dagger_{\mathbf{k}}]=1$ implies $u_{\mathbf{k}}^2-v_{\mathbf{k}}^2=1$ and we can choose symmetric and real coefficients $u_{\mathbf{k}}=u_{-{\mathbf{k}}}$, $v_{\mathbf{k}}=v_{-{\mathbf{k}}}$, yielding

	\begin{align}
H_0\approx&-\frac{z_N JN}{2}S(S+1) \\
&+ z_NJS\sum_\mathbf{k}\left(u_{\mathbf{k}}^2 + u_{\mathbf{k}}^2 + 2u_{\mathbf{k}}\textcolor{black}{v}_{\mathbf{k}}\gamma_{\mathbf{k}}\right)\left[\alpha^\dagger_\mathbf{k}\alpha_\mathbf{k}+\beta^\dagger_\mathbf{k}\beta_\mathbf{k}+1\right] \nonumber\\
& + z_NJS\sum_\mathbf{k}\left(\gamma_\mathbf{k}(u_\mathbf{k}^2+v_\mathbf{k}^2) + 2u_\mathbf{k}v_\mathbf{k}\right)\left[\alpha_{\mathbf{k}}\beta_{-\mathbf{k}} + \alpha^\dagger_{\mathbf{k}}\beta^\dagger_{-\mathbf{k}}\right]\nonumber
	\end{align}
	
\noindent The coefficients $u_{\mathbf{k}},v_{\mathbf{k}}$ of the transformation are now chosen such that the off-diagonal terms vanish:

	\begin{align}\label{offdiagzero}
\gamma_\mathbf{k}(u_\mathbf{k}^2+v_\mathbf{k}^2) + 2u_\mathbf{k}v_\mathbf{k}= 0,
	\end{align}

\noindent yielding the diagonal form

	\begin{align}
&H_0\approx-\frac{z_NJN}{2}S(S+1)+\sum_\mathbf{k}\hbar\omega_\mathbf{k}\left[\alpha^\dagger_\mathbf{k}\alpha_\mathbf{k}+\beta^\dagger_\mathbf{k}\beta_\mathbf{k}+1\right] \\
&\hbar\omega_\mathbf{k}\equiv z_NJS \left(u_\mathbf{k}^2+v_\mathbf{k}^2 + 2u_\mathbf{k}v_\mathbf{k}\gamma_\mathbf{k}\right)=z_NJS\sqrt{1-\gamma_\mathbf{k}^2}.\label{magnonfreq}
	\end{align}

\noindent Next, we similarly transform the light-induced perturbation $\delta H$. This produces equivalent results with the replacements

	\begin{align}
J\rightarrow \Delta J,\qquad \gamma_{\mathbf{k}}  \rightarrow   \xi_\mathbf{k}=\frac{1}{z}\sum_{\boldsymbol{\delta}}(\hat{e}\cdot\hat{\boldsymbol{\delta}})^2\exp(i\mathbf{k}\cdot\mathbf{\boldsymbol{\delta}}).
	\end{align}

\noindent Using Eq.~\eqref{offdiagzero} we write this as

	\begin{align}
 \delta H\approx& -\frac{z_N\Delta JN}{2}S(S+1)
+ \sum_\mathbf{k}\hbar\delta\omega_\mathbf{k}\left[\alpha^\dagger_\mathbf{k}\alpha_\mathbf{k}+\beta^\dagger_\mathbf{k}\beta_\mathbf{k}+1\right] + \\
 +& \hbar V_\mathbf{k}\left[\alpha_{\mathbf{k}}\beta_{-\mathbf{k}} + \alpha^\dagger_{\mathbf{k}}\beta^\dagger_{-\mathbf{k}}\right] \nonumber
	\end{align}

\noindent where
	\begin{align}
\hbar\delta\omega_\mathbf{k}= z_N\Delta JS \frac{\left(1-\gamma_\mathbf{k}\xi_\mathbf{k}\right)}{\sqrt{1-\gamma^2_\mathbf{k}}},\quad
\hbar V_\mathbf{k}= z_N\Delta J S \frac{\left(\xi_\mathbf{k} - \gamma_\mathbf{k}\right)}{\sqrt{1-\gamma^2_\mathbf{k}}}. 
	\end{align}
	
\noindent Hence we observe that in general $H_0$ and $\delta H_0$ cannot be simultaneously diagonalized. Therefore, under the presence light-induced perturbations to the exchange interactions, the terms $\alpha_{\mathbf{k}}\beta_{-\mathbf{k}} + \alpha^\dagger_{\mathbf{k}}\beta^\dagger_{-\mathbf{k}}$ remain which can annihilate or create magnon pairs with opposite $\mathbf{k}$. 

%We note that there is a subtle technical difference between the magnons described by the bose operators ak , bk and those described by ?k, ?k. The former represent excitation with respect to the collinear state (classical Ne?el state) and are strictly localized to one magnetic sublattice, while the latter describe excitations that are a superposition of magnons in different sublattices with respect to the ground state which is already dressed by two-magnon excitations. Mathematically, it is convenient to work with ? , ? , since they diagonalize H. However, it is equally possible to keep working in the basis ak , bk of magnons that are completely localized to either of the magnetic sublattices. Here we elaborate further on the difference between the two representations. 

\textcolor{black}{We note that there is a subtle technical difference between the magnons described by the Bose operators $a_\mathbf{k},b_\mathbf{k}$ and those described by $\alpha_\mathbf{k},\beta_\mathbf{k}$. The former represent excitation with respect to the collinear state (classical N\'eel state) and are strictly localized to one magnetic sublattice, while the latter describe excitations that are a superposition of magnons in different sublattices with respect to the ground state, which is already dressed by two-magnon excitations. Mathematically, it is convenient to work with $\alpha_\mathbf{k},\beta_\mathbf{k}$, since they diagonalize $H_0$. However, it is equally possible to keep working in the basis $a_\mathbf{k},b_\mathbf{k}$ of magnons that are completely localized to either of the magnetic sublattices. Here we elaborate further on the difference between the two representations. First of all, we stress that $\alpha_\mathbf{k}, \beta_\mathbf{k}$ almost coincide with  $a_\mathbf{k}$ and $b_\mathbf{k}$ when $\mathbf{k}$ is close to the BZ boundary where $v_\mathbf{k} \ll u_\mathbf{k}$ in Eq. (A5) and (A6). Hence, for many qualitative discussions, it is appropriate to think of 2M excitations as if they comprise a pair of magnons, one in each sublattice, each with the same frequency but with opposite $\mathbf{k}$ in the different sublattices. One such two-magnon state is depicted in Fig. 2b, while Fig. 2a shows the N\'eel state without any magnon excitation. It is, however, important to stress that the N\'eel state is not the exact eigenstate of the quantum antiferromagnet (see also the discussion in Sec. II.A.2). Therefore, even in the ground state, there are quantum fluctuations due to incoherent two-magnon excitations. Hence, already at zero temperature both the N\'eel state $|0\rangle |0\rangle$ and the state $a^{\dagger}_\mathbf{k} b^{\dagger}_\mathbf{-k} |0\rangle|0\rangle$ with two magnons excited have non-zero occupation, despite the fact that the energy separation between these states is $\Delta E\sim 2J$. A sudden perturbation $\Delta J$, therefore can induce coherent dynamics between the N\'eel state and the two-magnon excited state.}

\textcolor{black}{Using the transformed operators  $\alpha_\mathbf{k}, \beta_\mathbf{k}$ a conceptually different picture arises. Since the operators  $\alpha_\mathbf{k},\beta_\mathbf{k}$ diagonalize the Hamiltonian, the excited state $\alpha^{\dagger}_\mathbf{k} \beta^{\dagger}_\mathbf{-k} |0\rangle|0\rangle$ is not occupied in the ground state.  In this basis, the sudden perturbation of $J$ induces a coherence between the ground state $|0\rangle|0\rangle$   and excited state $\alpha^{\dagger}_\mathbf{k} \beta^{\dagger}_\mathbf{-k} |0\rangle|0\rangle$ that remains after the pulse, see the discussion in \cite{mentink2017jpcm}.
Of course, since the transformation between the two representations is unitary, both yield the same observables. For the qualitative discussion in Sec. II.A we mostly rely on the description using the $a_\mathbf{k},b_\mathbf{k}$ operators, since in this basis magnons are easier to visualize. On the other hand, for the theoretical derivations in Sec. II.B we use the magnons $\alpha_\mathbf{k}, \beta_\mathbf{k}$ since they are more convenient mathematically.}

\section{{Spin correlations and N\'eel vector}}\label{a:neelcorr}
Here we derive the expressions for spin correlations and N\'eel vector within harmonic magnon theory, using the same approach as introduced above. For the components of the N\'eel vector we obtain:

\begin{align}
 L^z &=NS-\sum_{i\in\Uparrow}  a_i^\dagger a_i + \sum_{j\in\Downarrow}b_{j}^\dagger b_{j}  \nonumber \\
&= NS - \sum_\mathbf{k}  a_\mathbf{k}^\dagger a_\mathbf{k} +  b_\mathbf{k}^\dagger b_\mathbf{k},\\
 L^x &=\sqrt{NS} \left( a_0+a_0^\dagger - (b_0+b_0^\dagger)\right), \nonumber \\
 L^y &=-i\sqrt{NS} \left( a_0-a_0^\dagger- (b_0-b_0^\dagger)\right) \label{Lxy},
\end{align}

Longitudinal spin correlations are defined as

\begin{align}
\sum_{\langle i,j\rangle} S_i^z S_j^z =-NzS^2/2+S\sum_{i,\boldsymbol\delta}  a_i^\dagger a_i + b_{i+\boldsymbol\delta}^\dagger b_{i+\boldsymbol\delta} \nonumber \\
= -Nz_NS^2/2 +z_NS \sum_\mathbf{k}  a_\mathbf{k}^\dagger a_\mathbf{k} +  b_\mathbf{k}^\dagger b_\mathbf{k}.
\end{align}

From this we observe the direct connection between $\langle L^z\rangle$ and $\sum_{\langle i,j\rangle}\langle S_i^z S_j^z\rangle$ (see also \cite{Bossini2016}):
\begin{align}
 L^z = \frac{NS}{2}-\frac{1}{z_NS}\sum_{\langle i,j\rangle} S_i^z S_j^z.
\end{align}
Subsitition of the Bogoliubov transformation gives for the longitudinal correlations

\begin{align}
&\sum_{\langle i,j\rangle} S_i^z S_j^z = -NzS^2/2 \nonumber \\
&\qquad\qquad+zS \sum_\mathbf{k} (u_\mathbf{k}^2+v_\mathbf{k}^2)\left( \alpha_\mathbf{k}^\dagger \alpha_\mathbf{k} +  \beta_{-\mathbf{k}}^\dagger \beta_{-\mathbf{k}}+1\right)\nonumber\\
&\qquad\qquad\qquad\qquad+2u_\mathbf{k} v_\mathbf{k}\left( \alpha_\mathbf{k} \beta_{-\mathbf{k}} +  \alpha_{\mathbf{k}}^\dagger \beta_{-\mathbf{k}}^\dagger\right)
\end{align}

\noindent Direct substitution of the magnon pair operators and using Eqs.~\eqref{offdiagzero}, \eqref{magnonfreq}, gives the formula for the longitudinal correlations Eq.~\eqref{eq_longitudinal_correlation} of the main text.

%\bibliography{MyRefs}

\end{document}